\UseRawInputEncoding
\documentclass[letterpaper,12pt,final]{article}

\usepackage{times}

\usepackage[breaklinks,colorlinks,linkcolor=black,citecolor=black,urlcolor=black]{hyperref}

\usepackage[margin=1in]{geometry}

\usepackage{graphicx}
\DeclareGraphicsExtensions{.pdf,.jpeg,.png}
\usepackage{cite}
\usepackage{amsmath}
\usepackage{titlesec}
\titlelabel{\thetitle.\quad}
\titleformat*{\section}{\normalfont\bfseries}
\titleformat*{\subsection}{\normalfont\bfseries}
\titleformat*{\subsubsection}{\normalfont\bfseries}
\titleformat*{\paragraph}{\normalfont\bfseries}
\titleformat*{\subparagraph}{\normalfont\bfseries}

\usepackage{amsmath, amsfonts, amssymb, amsthm, bm}

\usepackage{graphicx, subfigure} % if subfigure used, subfig could not be used, ??????????????in subfig, use \subfloat[]

\usepackage{multirow}
\usepackage{booktabs}
\usepackage{lineno}

\usepackage{color, soul}

\usepackage{appendix}

\newtheorem{myTheorem}{Theorem}
\newtheorem{myAssumption}{Assumption}
\newtheorem{myProof}{Proof}

\begin{document}
	\date{}
	
	\title{Formation Tracking for a Multi-Auv System Based on an Adaptive Sliding Mode Method in the Water Flow Environment}
	
	\author{% The following lines should be within the Author's block
		%\footnote{Manuscript received December 6, 2021, revised May 28, 2022, International Journal of Robotics and Automation}
		%Authors' information should be left commented for the submittal of the %manuscript.  The comments should be removed for the final submittal
		Xin Li\\
		%Department of Information Engineering\\
		Shanghai Maritime University\\
		%1550 Haigang Avenue\\
		%Shanghai, 201306, China, \\
		%email: lixin850224@163.com \\
		\and
		Daqi Zhu \\
		Unversity of Shanghai for Science and Technology\\
		\and
		Bing Sun \\
		Shanghai Maritime University\\
		\and
		Qi Chen \\
		Unversity of Shanghai for Science and Technology\\
		\and
		Wenyang Gan\\
		Shanghai Maritime University\\
		\and
		Zhigang Li\\
		Shenyang Institute of Automation\\
		%Department of Automation\\
		%University of XXXXXX\\
		%Address1\\
		%City, State, ZIP, Country \\
		%email: email@address.com \\
		%\and
		%FirstName LastName \\
		%Department of XXXXXXXXXX\\
		%University of XXXXXX\\
		%Address1\\
		%City, State, ZIP, Country \\
		%email: email@address.com \\
		}
	
	\maketitle
	
	\thispagestyle{empty}
	
	\noindent
	
	{\bf\normalsize Abstract}\newline
	{In this paper, formation tracking for a multi-AUV system (MAS) using an improved adaptive sliding mode control method is studied in the Three Dimensional (3-D) underwater environment.
		Firstly, the kinematics model and the dynamic model of the AUVs are given as the Six Dimensions of Freedom (6-DOF) considered.
		Then, control law based on the mathematical model of the AUVs is proposed based on the improved sliding mode method. A second order sliding mode control method is adopted to eliminate the chatting phenomenon of the controller.
		Thirdly, considering the water flow in the underwater working environment of the AUVs, an adaptive module is added to the controller. With the adaptive approach, the finite disturbances caused by water flow could be handled with the controller. The proposed method achieves stability by substituting an adaptive continuous term for the switching term in the controller.
		At last, a robust sliding mode controller with continuous model predictive control strategy for the multi-AUV system is developed to achieve leader-follower formation tracking under the presence of bounded flow disturbances, and simulations are implemented to confirm the effectiveness of the proposed method. } \vspace{2ex}
	
	\noindent
	{\bf\normalsize Key Words}\newline
	{Formation tracking, sliding mode control, adaptive control, leader-follower, water flow}
	
	\section{Introduction}
	
	Autonomous underwater vehicles (AUVs) are playing important roles in ocean explorations \mbox{\cite{Leonard2016,WOS:000654237000003}}, such as missions for seabed mapping, underwater surveillance, oil and gas exploration, salvage tasks, etc.
	As the underwater tasks become more and more complex, it is necessary to carry out underwater missions using multiple AUV systems (MASs). In certain missions, the AUVs should move collectively as a formation.
	Formation control is a technology to control a group of vehicles, including ground robots, aerial crafts, surface vehicles and AUVs moving along the desired path as the task requires while maintaining desired formation patterns and adapting to the environmental constraints, such as obstacles, ocean currents, and limited space \mbox{\cite{CUI20101491,WOS:000792022100007}}. In recent years, a number of scholars have investigated the formation control problem as a key technology for MAS control. From a general viewpoint, a qualified formation controller must realize MAS's instant trajectory tracking and the stability of the formation shape while facing unknown bounded disturbances, such as ocean currents and communication delays.
	
	According to recent literature, consensus control method has been proposed for MASs with a centralized structure or a distributed structure \cite{renconsensus2010,zhu_survey_2017}. Consensus control for MAS formation can be used with or without a formation leader \cite{renConsensus2007,CAO20091299}, and it is often used to handle the situation of time-delays in the MAS. However, most of the researches based on consensus controller are focused on the mathematical derivation and stability proof, without the practical system's dynamics taken into consideration. As a result, the consensus control method can not be used to implement path planning and formation tracking for the MAS in complex environments \cite{tianconsensus2008,linfurther2009}.
	Rigid body concept is often used in the formation keeping researches, as well as the digraph or undigraph theories. Based on these methods and event-triggered concept, great formation control type can be achieved without dynamic analysis \cite{wen_asynchronous_2018}. In the literature \cite{cai_adaptive_2015}, dynamics of the formation system are considered, but analysis is focused on the structure layer, not the implementation layer.
	In most existing works, the agent dynamics are restricted to be first-order integrators, and the proposed consensus protocols are based on relative states among neighboring agents, which in many cases are not available \cite{wangconsensus2008,lidaynamic2011}. In the work of \cite{GIULIETTI200565}, general linear or linearized dynamics of agents are studied for formation flight, which is similar to the MAS formation, and communication topology contains a directed spanning tree is proposed, but the technique in this study lacks efficiency to realize the dynamic control of the agents in the formation.
	In the research of \cite{yang_leaderfollower_2018}, the output synchronization of leader-follower systems with an active leader is formulated as a distributed optimal tracking problem, and inhomogeneous algebraic Riccati equations are derived to solve it. This research is using the most popular machine learning method to solve the MAS control problem, which is innovative and promising. However, in order to solve the disturbance problem in the MAS formation control, such as the water flow influences, traditional control theories seems to be more powerful.
	
	To keep the AUVs' formation on the right path, the system must be stable and the controller must be robust. To achieve this purpose, the AUV's dynamics are inevitably taken into the controller \cite{Wenyang}.
	As the MASs are nonholonomic and underactuated, it is difficult to build the system's accurate model and realize dynamic control.
	In recent years, some scholars are trying to do researches about formation control regarding dynamics.
	The commonly used dynamic control techniques are PID, LQR, Feedback Linearization, and Sliding Mode Control, etc.
	Those techniques have achieved the formation control purpose to some extent, while most of them ignore the presence of environmental disturbances or model uncertainties \cite{WOS:000459467100007}. A control method based on an enhanced reduced-order extended state observer is proposed to control the motor-wheels dynamic	model of a differential driven mobile robot to deal with single system's model uncertainties and external disturbances, which could be further employed with multiagent systems \cite{9629270, https://doi.org/10.1049/pel2.12392}.
	Another well-known dynamic control technique is sliding mode control. It has certain advantages such as the insensitivity to parameter variations to achieve robustness and guarantee the system stability, but it presents the inconvenience of high-frequency switching of the control signal \cite{Fahimisliding2007,WOS:000459467100004}.
	Recent researches in sliding mode control theory present a kind of robust, continuous, and even smooth controller suitable for MAS formation \cite{Shtessel2008secondorder, Edwards1998}. Some additional methods for SMC have been proposed to reduce the system's chattering. For example, adaptive sliding mode control methods have been applied to underwater vehicles \cite{SOYLU20081647, 1186752}. However, the disturbances caused by the water flow in the AUV working environments were not taken into consideration \cite{5968958, WOS:000719381800001}.
	The most important advantage of the adaptive method is that it allows the development of the controllers to be robust, accurate and continuous for each controlled plant, so that the controllers could resist finite disturbances. At the same time, higher-order sliding mode control technique preserves the properties of standard SMC and removes the chattering effects \cite{1283410,1532406}. However, external disturbances are usually not considered in the controller. For the multi-AUV formation control, an adaptive sliding mode control method has been proposed in the work of \cite{li_formation_2016}, handling the chattering problem and distubances, and simulations have been done to verify its effectiveness. However, the model of the water flow was relatively sketchy and inconsistent with reality. \cite{8661753} presents a fuzzy neural network controller using impedance learning for coordinated multiple constrained robots to deal with the presence of the unknown robotic dynamics and the unknown environment. Unfortunately, oscillations controller has not been designed and verified. \cite{WANG2020101971,MENG2021102781} address the leader-follower formation control for multi-AUV systems based on improved sliding mode controllers, without considering the actual mechanical structure of AUVs. What's more, simulations on the software level can not fully reflect the AUVs' dynamic characteristics.
	
	To deal with the formation control and formation tracking problems, we adopt the improved SMC approach to replace the discontinuous term of the conventional sliding mode controller with an estimate of the uncertainties with model predictive control in an adaptive way. Simulations of the AUV formation control problem in the water flow environment has been done using MATLAB and Gazebo with physics engine to confirm the effectiveness of the proposed method. The combination of the SMC method and adaptive method could make better results for the MAS formation control, which is further studied in this paper.
	
	The organization of the paper is as follows. A hierarchical controller structure is proposed for MAS formation control in Section II. The AUVs' kinematics and dynamics are represented in the specific reference frames in section III, where the water flow models are established as well. In Section IV, we introduce the adaptive sliding mode control method and set the propeller configuration for simulations. In section V, simulation results are presented for formation control of a group of AUVs in the 3-D environment. Some conclusive remarks are given in Section VI.
	
	\section{Hierarchical Controller Structure for MAS Formation}
	
	Aiming at the particularity of underwater environment and the requirement of AUV underwater tasks, this paper designs a hierarchical controller structure for MAS formation.
	
\begin{figure}[tbh]
		\centering
		\includegraphics[width=0.7\textwidth]{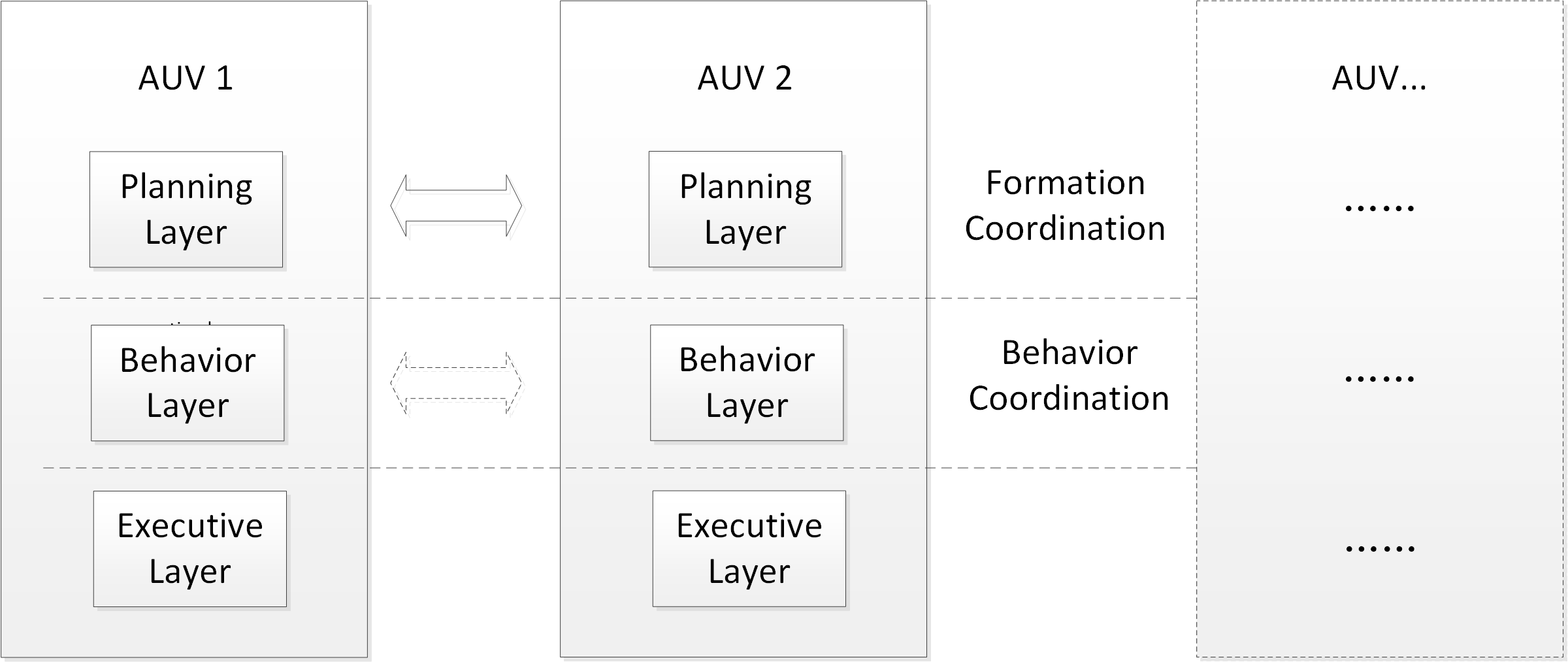}%
		\caption{Multi-AUV hierarchical control system design.}
		\label{fig_1}
	\end{figure}

\begin{figure}[tbh]
		\centering
		\includegraphics[width=0.8\textwidth]{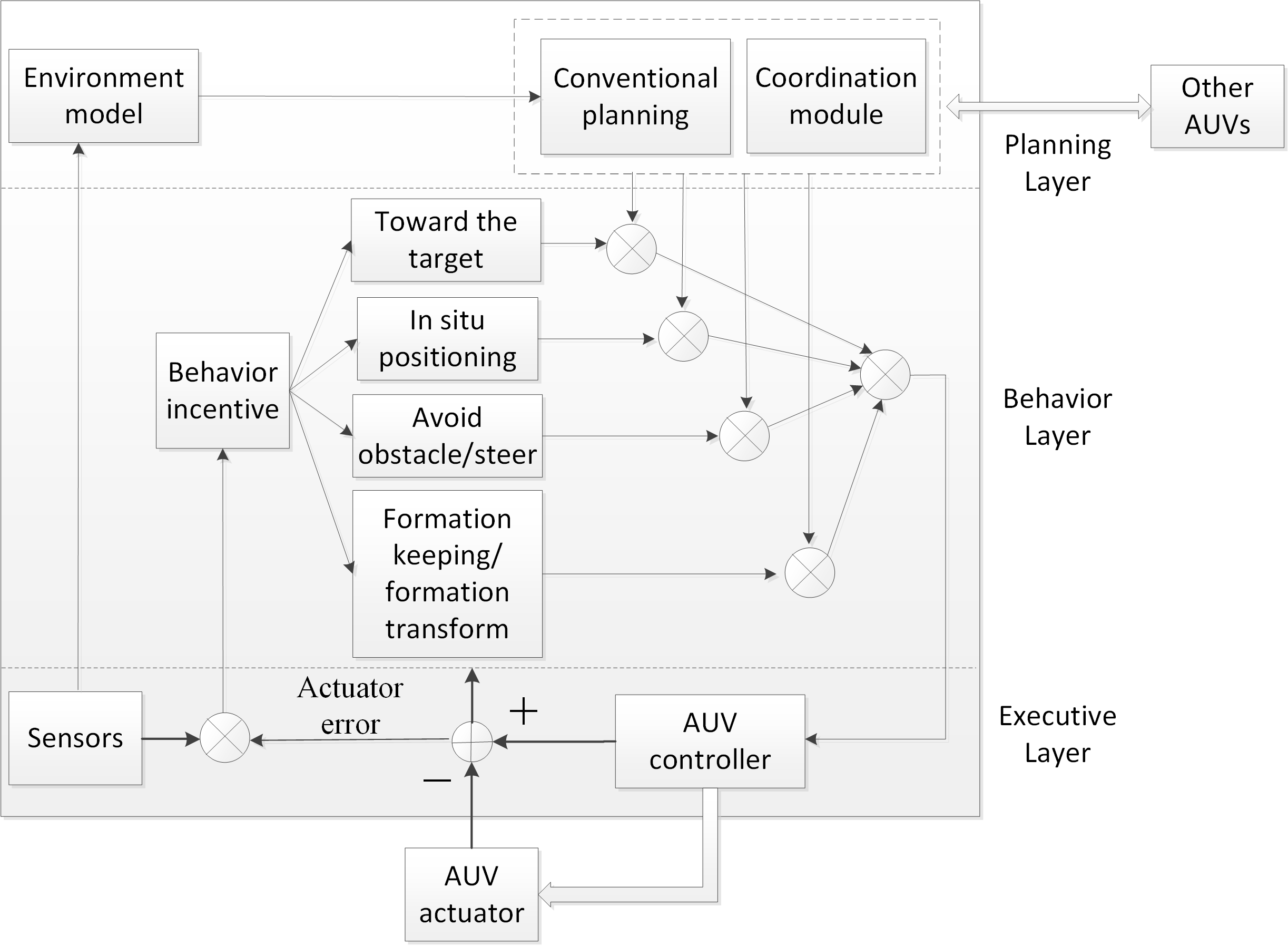}%
		\caption{The AUV's hybrid control system architecture.}
		\label{fig_2}
	\end{figure}

	As shown in \mbox{Fig. \ref{fig_1}}, an AUV's control system is a constitutional unit of the whole formation of the MAS. An open formation control system is built based on the basic single unit of each AUV. Note that each AUV has a hybrid control system, which is shown in \mbox{Fig. \ref{fig_2}}. This kind of formation controller consists of three levels, namely the planning layer, the behavior layer, and the executive layer. The three layers are top-down designed to make the controller easy to realize. At the same time, the formation controller is open and extensible, benefiting from the clear hierarchy.
	
	Fig. \ref{fig_1} and Fig. \ref{fig_2} illustrate the design concept of the formation controller.
	In the top layer, i.e. the planning layer, formation strategies including formation keeping, formation tracking and task assignment are set with kinematic control algorithms as the core algorithms of this layer. As this layer is for the strategy and kinematic control, it is top and general.
	The behavior layer mainly realizes the behavioral coordination, including specific actions of the AUV such as heading to the target, localizing in the position, and avoiding obstacles, etc. Obstacle avoidance algorithms and optimal path planning methods are the core algorithms of the behavior layer.
	In the executive layer, dynamic controller is implemented, which provides anti-interference ability and robustness for the whole formation controller. The three levels of the formation control system are designed from top to bottom and complement each other, but also have some mutual infiltration.
	For example, the planning layer also provides some behavior actions, such as formation transformation, collision avoidance among members in the formation, etc.
	The implementation of specific actions mainly depends on the behavior layer, while the control of dynamics is mainly realized by the execution layer.
	As mentioned in the former section, a lot of related literature ignore the dynamics of the MAS, this paper mainly focus on the design of the executive layer in the bottom of the controller's structure.
	
	%This is how to insert a Figure - uncomment when using.
	%Figure file should be in the same directory as manuscript
	
	%\begin{figure}[!t]
	%\centering
	%\includegraphics[width=4 in]{Diagram1.png}
	%\caption{This is the Caption}
	%\label{fig:FigureLabel}
	%\end{figure}
	
	%Equations are automatically numbered
	%\begin{equation}
	%f(x) = \alpha \sin (\omega t + \beta) + \gamma
	%\label{eq_model}
	%\end{equation}

	\section{The AUV's Mathematical Model}
	
	This paper focuses on AUVs' dynamic control, and the proposed method aims at the optimization of the control procedure. In the beginning, it's necessary to investigate the kinematics of an AUV, because all the efforts put on dynamic control is to obtain effective kinematic control results, such as the velocity and trajectory of the AUV.
	
	In the design of the AUV's controller, water flow caused disturbances can be usually expressed as constants or slowly changing quantities. It is difficult to find a unified mathematical model because the mode of ocean current is time-varying. Therefore, the disturbances are treated as bounded variables and a certain flow function is constructed in this paper to verify the effectiveness of the algorithm.
	
	\subsection{Kinematics of an AUV}
	
\begin{figure}[tbh]
		\centering
		\includegraphics[width=0.5\textwidth]{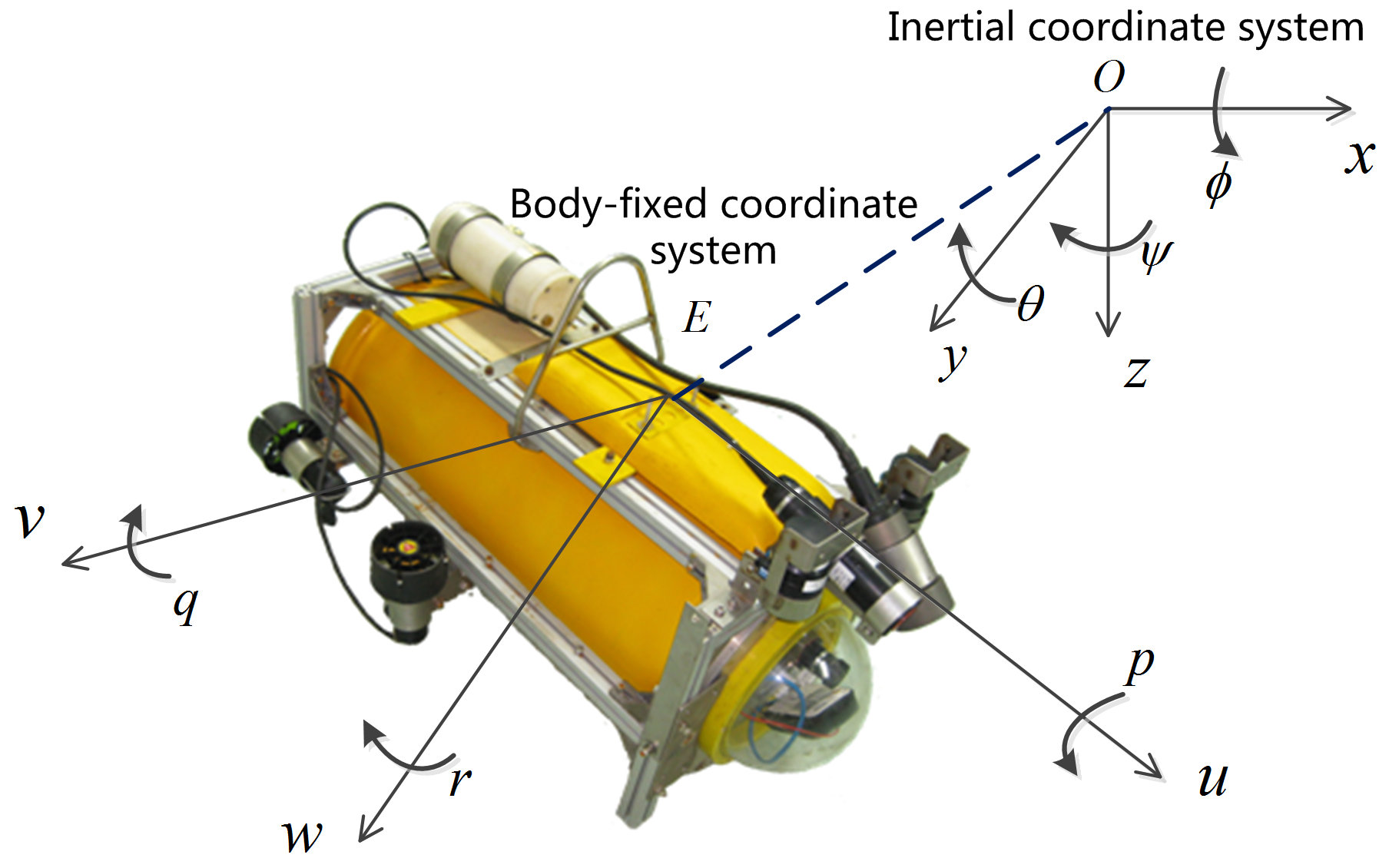}%
		\caption{The AUV's coordinate systems.}
		\label{fig_3}
	\end{figure}

	As shown in Fig. \ref{fig_3}, the state vector (position and posture) of the AUV is defined as $ \boldsymbol{\eta} = \left[\boldsymbol{\eta}_{1}^{\mathrm{T}} \ \boldsymbol{\eta}_{2}^{\mathrm{T}}\right]^\mathrm{T} \in \mathbb{R}^{6}$, where $\bm{\eta}_{1}=\left[x \  {y} \  {z}\right]^{\mathrm{T}} \in \mathbb{R}^{3}$ is the vector of vehicle position coordinates in an earth-fixed or inertial referenced frame and $\bm{\eta}_{2}=\left[\phi \  {\theta} \  {\psi}\right]^{\mathrm{T}} \in \mathbb{R}^{3}$ is the vector of vehicle's Euler-angle coordinates in an inertial reference frame. At the same time, the dynamic vector of the AUV is defined as $ \boldsymbol{v} = \left[\boldsymbol{v}_{1}^{\mathrm{T}} \ \boldsymbol{v}_{2}^{\mathrm{T}}\right]^\mathrm{T} \in \mathbb{R}^{6}$, where $\bm{v}_{1}=\left[u \  v \  w\right]^{\mathrm{T}} \in \mathbb{R}^{3}$ is the vector of vehicle linear velocity expressed in the vehicle-fixed reference frame, and $\bm{v}_{2}=\left[p \  q \  r\right]^{\mathrm{T}} \in \mathbb{R}^{3}$ is the vector of vehicle angular velocity expressed in the body-fixed frame.
	
	%\begin{myRemark}
	%	Two kinds of coordinate systems are defined in this paper: the inertial reference frame represented as $\sum_{i} :\{O-x y z\}$, and the vehicle-fixed reference frame denoted by $\sum_{v} :\left\{E-x_{v} y_{v} z_{v}\right\}$. The $z$ axis is parallel to the direction of gravity.
	%	The positive direction of the axis $x_v$ is the same to the positive surge direction of AUV. When the AUV is floating on the water surface, axis $z_x$ is parallel to axis $z$.
	%	All the coordinates follow the right hand rule.
	%	\label{remark_1}
	%\end{myRemark}
	
	The parameters of the AUV's 6-DOF model is show in Table \ref{table_1}. The vehicle-fixed linear and angular velocities and the time derivative of the earth-fixed vehicle coordinates are related by the following equations:
	
	\begin{equation}
		\label{eqn_1}
		\left\{ \begin{array}{ll}
			\boldsymbol{v}_{1}=\boldsymbol{R}_{I}^{C} \dot{\eta}_{1} \\
			\bm{v}_{2}=\boldsymbol{T} \dot{\eta}_{2}
		\end{array} \right.
	\end{equation}
	where $\bm{R}_{I}^{C}$ is the rotation matrix expressing the linear velocity transformation from the inertial earth-fixed frame to the body-fixed frame, and the  angular velocity transformation matrix $\bm{T} \in \mathbb{R}^{6}$ is Jacobian matrix, which can be both found in former literatures \cite{Antonelli2001,Fossen1994,Antonelli2014}.
	
	% \begin{equation}
		%     \label{eqn_2}
		%     \bm{v}_{2}=\boldsymbol{T} \dot{\eta}_{2}
		% \end{equation}
	
	\begin{table}[htb]
		\caption{The 6 DOF parameters of the AUV}
		\centering
		\label{table_1}
		\begin{tabular}{l c c c}
			\hline
			Kinematics &  Force/Moment & Velocity/Angular velocity & Position/Angel \\
			\hline
			Surge & $X$ & $u$ & $x$ \\
			Sway & $Y$ & $v$ & $y$ \\
			Heave & $Z$ & $w$ & $z$ \\
			Roll & $K$ & $p$ & $\phi$ \\
			Pitch & $M$ & $q$ & $\theta$ \\
			Yaw & $N$ & $r$ & $\psi$ \\
			\hline
		\end{tabular}
	\end{table}
	
	\subsection{Dynamics of the AUV}
	
	Considering the coordinate systems illustrated in Fig. \ref{fig_3}, the dynamic equations of the AUVs in the vehicle-fixed reference frame are written as
	
	\begin{equation}
		\label{eqn_2}
		\left\{ \begin{array}{ll}
			\mathbf{M}(\bm{q}) \dot{\bm{q}}+\mathbf{C}(\boldsymbol{q}) \boldsymbol{q}+\mathbf{D}(\boldsymbol{q}) \boldsymbol{q}+\mathbf{g}(\boldsymbol{e})=\boldsymbol{\tau} - \boldsymbol{\tau}_c \\
			\dot{\bm{e}}=\mathbf{J}(\bm{e})\cdot\bm{q}
		\end{array} \right.
	\end{equation}
	where $\bm{q}=\left[u\ v\ w\ p\ q\ r\ \right]^{\mathrm{T}}$ is the ROV spatial velocity state vector with respect to its body-fixed reference frame, and $\bm{e}=\left[x\ y\ z\ \phi\ \theta\ \psi \right]^{\mathrm{T}}$ is the position and orientation state vector with respect to the inertial reference frame.
	The spatial transformation matrix between the inertial frame and the AUV's body-fixed frame can be defined through the Euler angle transformation, denoted by $\mathbf{J}(\bm{e}) \in \mathbb{R}^{6 \times 6}$.
	The term $\mathbf{M}(\bm{q}) \in \mathbb{R}^{6 \times 6}$  is the inertia matrix including the added mass effects.
	$\mathbf{C}(\boldsymbol{q}) \in \mathbb{R}^{6 \times 6}$ is the matrix of centrifugal and Coriolis terms.
	$\mathbf{D}(\boldsymbol{q}) \in \mathbb{R}^{6 \times 6}$ is the drag matrix, with that $\mathbf{g}(\boldsymbol{e}) \in \mathbb{R}^{6 \times 6}$ is the vector of gravity and buoyancy forces and moments.
	$\boldsymbol{\tau} \in \mathbb{R}^{6 \times 1}$ is the matrix of control forces and moments acting on the AUV center of mass \cite{SOYLU20081647,5229987,Chassignet2009}. $\boldsymbol{\tau}_c \in \mathbb{R}^{6 \times 1}$ is the matrix of forces and torques produced by the water flow which is acting on the AUV working under the water. Note that as the disturbance is caused by the water flow, $\boldsymbol{\tau}_c$ is complex and changeable, so that it is difficult to build a model.
	
	%\begin{myRemark}
	%	\label{remark_2}
	%	In the design of the AUV's controller, water flow caused disturbances can be usually expressed as constants or slowly changing quantities. It is difficult to find a unified mathematical model because the mode of ocean current is time-varying. Therefore, the disturbances are treated as bounded variables and a certain flow function is constructed in this paper to verify the effectiveness of the algorithm.
	%\end{myRemark}
	
	The dynamic equation of motion of AUVs in the inertial reference frame can be presented as
	
	\begin{equation}
		\label{eqn_3}
		\bm{f}=\mathbf{M}_{\bm{e}}(\bm{e}) \ddot{\bm{e}}+\mathbf{C}_{\bm{e}}(\bm{q}, \bm{e}) \dot{\bm{e}}+\mathbf{D}_{\bm{e}}(\bm{q}, \bm{e}) \dot{\bm{e}}+\mathbf{g}_{\bm{e}}(\bm{e})=\mathbf{J}^{-\mathrm{T}} (\boldsymbol{\tau} - \boldsymbol{\tau}_c)
	\end{equation}
	where $\mathbf{M}_{e}(\bm{e})=\mathbf{J}^{-\mathrm{T}} \mathbf{M} \mathbf{J}^{-1}$, $\mathbf{C}_{e}(\boldsymbol{q}, \boldsymbol{e})=\mathbf{J}^{-\mathrm{T}}\left[\mathbf{C}-\mathbf{M} \mathbf{J}^{-1} \dot{\mathbf{J}}\right] \mathbf{J}^{-1}$, $\mathbf{D}_{e}(\boldsymbol{q}, \boldsymbol{e})=\mathbf{J}^{-\mathrm{T}} \mathbf{D} \mathbf{J}^{-1}$, and $\mathbf{g}_{e}(e)=\mathbf{J}^{-\mathrm{T}} \mathbf{g}$. The system dynamics are not exactly known, because the AUV dynamics are underactuated and dominated by hydrodynamic loads. It is difficult to accurately measure or estimate the hydrodynamic coefficients that are valid for AUVs' operating conditions. Therefore, the system dynamics can be written as the sum of estimated dynamics $\hat{\boldsymbol{f}}$ and the unknown dynamics $\tilde{\boldsymbol{f}}$ and we have
	
	\begin{equation}
		\label{eqn_4}
		\boldsymbol{f}=\hat{\boldsymbol{f}}+\tilde{\boldsymbol{f}}.
	\end{equation}
	
	The estimated dynamics vector is defined as
	
	\begin{equation}
		\label{eqn_5}
		\hat{\boldsymbol{f}}=\hat{\mathbf{M}}_{e}(\boldsymbol{e}) \ddot{\boldsymbol{e}}+\hat{\boldsymbol{h}}(\boldsymbol{q}, \boldsymbol{e})
	\end{equation}
	
	with $\hat{\bm{h}}(\boldsymbol{q}, \boldsymbol{e})=\hat{\mathbf{C}}_{e}(\boldsymbol{q}, \boldsymbol{e}) \boldsymbol{e}+\hat{\mathbf{D}}_{e}(\boldsymbol{q}, \boldsymbol{e}) \boldsymbol{e}+\hat{\mathbf{g}}_{e}(\boldsymbol{e})$ and the unknown dynamics vector are defined as
	\begin{equation}
		\label{eqn_6}
		\tilde{\boldsymbol{f}}=\tilde{\mathbf{M}}_{e}(\bm {e}) \ddot{\bm{e}}+\tilde{\boldsymbol{h}}(\boldsymbol{q}, \bm{e})
	\end{equation}
	with $\tilde{\boldsymbol{h}}(\boldsymbol{q}, \boldsymbol{e})=\tilde{\mathbf{C}}_{e}(\boldsymbol{q}, \boldsymbol{e}) \boldsymbol{e}+\tilde{\mathbf{D}}_{e}(\boldsymbol{q}, \boldsymbol{e}) \boldsymbol{e}+\tilde{\mathbf{g}}_{e}(\boldsymbol{e})$.
	$\tilde{\mathbf{M}}_{e}$, $\tilde{\mathbf{C}}_{e}$, $\tilde{\mathbf{D}}_{e}$, $\tilde{\mathbf{g}}_{e}$ represents the unknown items.
	In (\ref{eqn_6}), as the additional disturbance is bounded in the environment, the nonlinear uncertainty vector $\tilde{\bm{f}}$ and its time domain differential could be assumed to be bounded.
	
	\subsection{Modelling the water flow}
	
	In the previous researches on multi-AUV systems, there are relatively few researches on the ocean current. However, in practice, the influence of the current cannot be ignored, and it is even a very important factor for the successful tasks of the AUV formation.
	The existence of the water flow can generate drifting motions, making AUVs deviate from the predetermined trajectories.

	In the inertial coordinate system, the influence of the regular current can be expressed as the vector $\boldsymbol{\tau}_{e,f}$ of the force and torque. $\tilde{\boldsymbol{\tau}}_{e,f}$ is the force and torque vector generated by the propeller to offset the disturbance of ocean current, then we have $\tilde{\boldsymbol{\tau}}_{e,f} = - \boldsymbol{\tau}_{e,f}$ in the ideal condition. It is either a constant or as a quantity that transforms slowly and regularly.
	For the wind currents and other flows, the speeds change as time varying, generally after the wind stops the flow can continue change for a period. Such currents' information are hard to predict or measure.
	In the inertial coordinate system, it is assumed that the change of current disturbance is $\bm{d}_o \in \mathbb{R}^{6 \times 1}$ equivalent to $\bm{d}_{e,o}$ in the body-fixed frame. If $\bm{d}_{e,o}$ is bounded, adaptive higher-order controller adopted in this paper can deal with the disturbances in a certain range and ensure the normal operation of AUVs in formation.
	Considering the influence of regular current and variable flow, the total influence on the AUVs is
	\begin{equation}
		\label{eqn_7}
		\boldsymbol{\tau}_{e,c} = \boldsymbol{\tau}_{e,f} + \bm{d}_{e,o}
	\end{equation}
	where $\boldsymbol{\tau}_{e,c} = \mathbf{J}^{-\mathrm{T}} \boldsymbol{\tau}_c$ as depicted in (\ref{eqn_3}).
	
	Since the pattern of ocean current is time-varying, it is difficult to find a unified mathematical model, and the general disturbance of ocean current is bounded. Therefore, we build a function to verify the effectiveness of the algorithm by means of a trial water flow example \cite{1315730}.
	
	The 3-D working environment of AUVs is represented by cartesian coordinate system. Let $z=0$ be the water surface level, and the $z$-axis points to the bottom of the sea or river. The working space is stratified by depth $z$ into $N$ ($N \in \mathbb{N}^{+}$) layers, and each layer is an $x$-$O$-$y$ plane of two dimensions (2-D). Each $z=N$ ($N=1,2,3,...$) plane is rasterized, and each grid is a square with equal side length of $1$ ($m$). The currents in the areas within each grid can be regarded as the same. An east-west flow field with a meandering north-south flow field is adopted, and the mathematical expression of the flow function $C(x,y,t)$ at time $t$ is
	
	\begin{equation}
		\label{eqn_8}
		C(x, y, t)=1-\tanh \frac{y-B(t) \cos (k(x-c t))}{\sqrt{1+k^{2} B^{2}(t) \sin ^{2}(k(x-c t))}}.
	\end{equation}
	
	In (\ref{eqn_8}), $B(t)=\left(B_{0}+E \cos (\omega t+\theta)\right)$, $B_0=1.2$, $w=0.4$, $\theta = \pi/2$, $c=0.12$, $k=0.82$, $E=0.3$. According to the flow function, the current velocity can be calculated as
	
	\begin{equation}
		\label{eqn_9}
		\left. \begin{array}{ll}
			U(x, y, t)=-\frac{\partial C(x, y, t)}{\partial y} \\
			V(x, y, t)=\frac{\partial C(x, y, t)}{\partial x}
		\end{array} \right\}
	\end{equation}
	where $U(x, y, t)$ and $V(x, y, t)$ are the velocity components of the water flow in the $x$ and the $y$ directions at time $t$, respectively.
	
\begin{figure}[htb]
		\centering
		\subfigure[]{\includegraphics[width=0.6\textwidth]{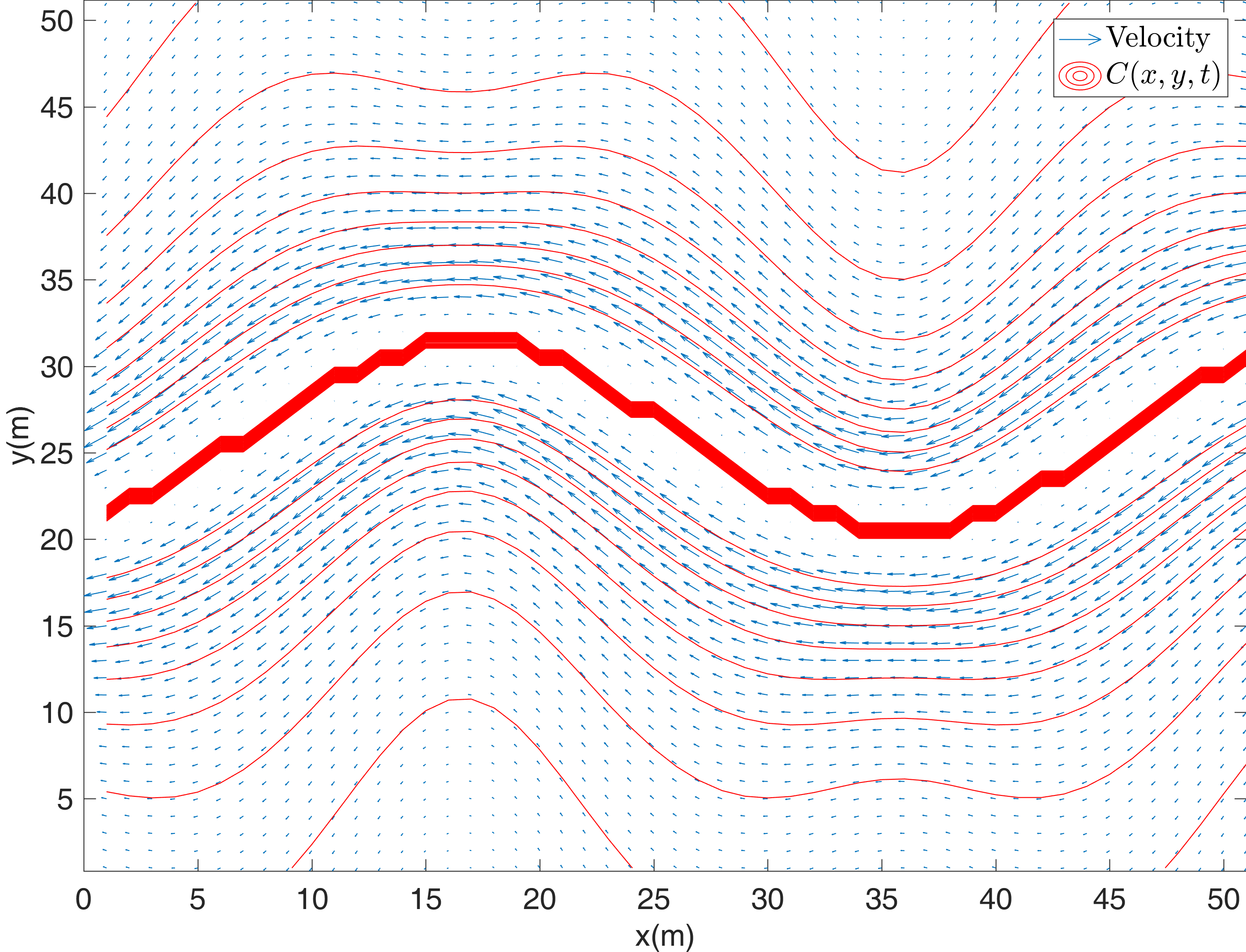} \label{fig_4a}}
		\\
		\subfigure[]{\includegraphics[width=0.6\textwidth]{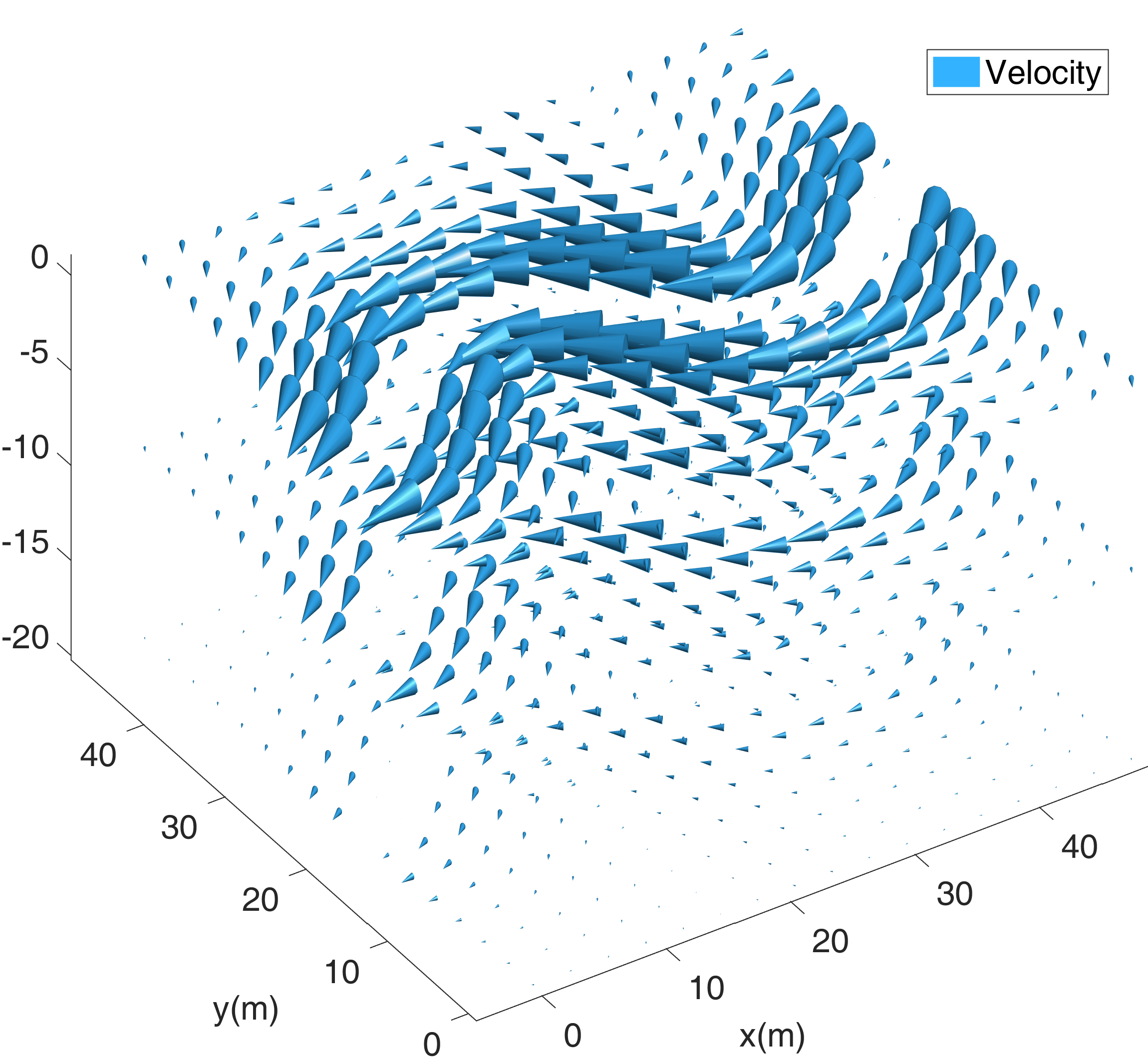} \label{fig_4b}}
		\caption{2-D and 3-D flow field generated by the water flow function. (a) 2-D workspace, (b) 3-D workspace.}
		\label{fig_4}
	\end{figure}

	Suppose that the water flow field in the workspace of a depth of $z \in [-20 \ 0]$ ($m$) is equally divided into three layers, and the velocity gradually decreases from the water surface to the bottom layer. $\left|V_{c i}\right|$ ($i=1,2,3$) denotes the the module value of the current velocity in each layer, with $\left|V_{c 1}\right|=2.4\left|V_{c 2}\right|=4\left|V_{c 3}\right|$. Then the 2-D current field is illustrated in Fig. \ref{fig_4a} and the 3-D current field is illustrated in Fig. \ref{fig_4b}.
	As shown in Fig. \ref{fig_4}, the arrow indicates the direction of the current at this position, and the length of the arrow is proportional to the velocity of the current. Taking this kind of water flow field as an example, simulations to test the proposed algorithm can be implemented.
	
	\section{The Improved Sliding Mode Controller}
	
	An adaptive sliding mode controller is proposed and implemented for multi-AUV systems. The AUVs could form a formation as needed and implement formation tracking. As a rule, sliding mode control could be divided into two parts. Firstly, we define the sliding manifold. Secondly, we find a control law to move the system trajectory towards the sliding manifold with finite time. In order to reduce the chattering phenomenon, we adopt a kind of higher-order sliding mode control law and add the adaptive features to it.
	
	\subsection{Sliding mode for the AUVs}
	
	The sliding order $r$ refers to the number of zeros of the continuous full derivatives which equal to zero on the sliding mode surface of the sliding mode variable $s$.
	The sliding set of order $r$ related to the sliding mode surface is defined as the equation $s=\dot{s}=\ddot{s}=\ldots=s^{(r-1)}=0$.
	If the sliding set of order $r$ exists, and it is assumed to be the local set in the sense of Filippov, the motion satisfying this equation is named $r$ order sliding mode about the sliding mode surface $s(x,t)=0$, and the order $r-1$ is called the relative order of the system.
	
	The sliding order of the AUV system is designed as a number of continuous total derivatives of in the vicinity of the sliding mode. It fixes the dynamics smoothness degree. The $r$-th order sliding mode is determined by the equation:
	
	\begin{equation}
		\label{eqn_10}
		\sigma=\dot{\sigma}=\ddot{\sigma}=\cdots=\sigma^{(r-1)}.
	\end{equation}
	
	The main problem of higher-order sliding mode control is the increment of the demanded information. For example, the system degree is 3, and then a 3-sliding controller needs an input control parameters $\sigma=\dot{\sigma}=\ddot{\sigma}=\sigma^{(r-1)}$. Fortunately, the super twisting algorithm that we adopt needs only the measurement of $\sigma$. \textcolor{red}{Then the  higher-order controller's action $u_{\textrm{HOSMC}}$ can be defined as}
	
	\begin{equation}
		\label{eqn_11}
		u_{\textrm{HOSMC}} = u_1 + u_2.
	\end{equation}
	
	Note that $u_1$ and $u_2$ are both functions of time $t$, corresponding respectively to a continuous function of the sliding mode surface and an integration of the sliding mode surface in the time domain \cite{Levant2003Higher}:
	
	\begin{equation}
		\label{eqn_12}
		u_{1}=\left\{\begin{array}{l}{-\lambda\left|\sigma_{0}\right|^{\rho} \operatorname{sign}(\sigma),|\sigma|>\sigma_{0}} \\ {-\lambda|\sigma|^{\rho} \operatorname{sign}(\sigma),|\sigma| \leq \sigma_{0}}\end{array}\right.
	\end{equation}
	
	and
	
	\begin{equation}
		\label{eqn_13}
		\dot{u}_{2}=\left\{\begin{array}{ll}{-u,} & {|u|>u_{\max }} \\ {-W \operatorname{sign}(\sigma),} & {|u| \leq u_{\max }}\end{array}\right.
	\end{equation}
	where $u_{max} \in N^+$, and can be normalize to 1. The remaining parameters are set in the simulations. According to \cite{Pisano2011Sliding}, the sufficient conditions to ensure convergence to the origin for the sliding mode plane in finite time is
	
	\begin{equation}
		\label{eqn_14}
		\left. \begin{array}{l}
			W>\frac{\Phi}{\Gamma_{M}} \\
			0<\rho \leq 0.5 \\
			\lambda^{2} \geq \frac{4 \Phi \Gamma_{M}(W+\Phi)}{\Gamma_{m}^{2}(W-\Phi)}
		\end{array}
		\right\}
	\end{equation}
	where $W$, $\rho$, $\lambda$, $\Phi$ are positive constants to be adjusted in the simulation.
	
	The algorithm does not need any differential information of sliding mode surface $\sigma$ in time domain, so the computation burden of controller is greatly reduced.
	Although the algorithm has good robustness, when $\sigma$ is very small, the control signal output is not Lipshitz, which may cause some noise in control output. In the simulations, parameters of the controller are continuously adjusted to find the most suitable values.
	
	On the phase plane defined by the estimable second-order state equation (\ref{eqn_5}), a first order dynamic equation is designed to represent the switching surface $\sigma$:
	
	\begin{equation}
		\label{eqn_15}
		\sigma=\dot{\boldsymbol{\varepsilon}}+2 \mathbf{M} \boldsymbol{\varepsilon}+\mathbf{M}^{2} \int \boldsymbol{\varepsilon} dt
	\end{equation}
	
	where $\bm{M} \in \mathbb{R}^{6}$ is positive definite diagonal matrix for the expected error response. When $\sigma = 0$, the dynamic response of the system is as expected, which means $\sigma$ can represent the difference between the current system state and the desired state. $\boldsymbol{\varepsilon}=\boldsymbol{e}-\boldsymbol{e}_{d}$ represents the tracking error between the actual system and the desired system. $\boldsymbol{e}_d$ is the desired AUV position and attitude vector. Substituting $\bm{\varepsilon}$ in (\ref{eqn_15}), we have
	
	\begin{equation}
		\label{eqn_16}
		\sigma=\dot{\boldsymbol{e}}-\dot{\boldsymbol{e}}_{r}
	\end{equation}
	where $\dot{\boldsymbol{e}}_{r}=\dot{\boldsymbol{e}}_{d}-2 \mathbf{M} \bm{\varepsilon} -\mathbf{M}^{2} \int \bm{\varepsilon} d t$, and $\bm{r}_r$ represents the reference trajectory of the AUV.
	
	As shown in (\ref{eqn_11}), $u_{\textrm{HOSMC}}$ represents the control force acting on the AUV's center of mass; $u_1$ and $u_2$ represent the equivalent control and switching control respectively.
	The equivalent control $u_1$ is a continuous function based on the AUV's model. If there is no uncertainty in the system dynamics, the equivalent control rate can achieve the desired state.
	However, AUV has a dynamic model and works with external disturbances, so the controller should include a switching control quantity $W \operatorname{sign}(\sigma)$, where $W$ is the upper limit of system model parameter uncertainty.
	the switching control $u_2$ could be a discontinuous feedback function, which is mainly used to compensate the differential characteristics and external disturbances of the expected AUV's dynamics, so as to make the system stable.
	However, the switching control rate function is easy to make the system oscillate near the switching surface, causing the chattering phenomenon, so it is necessary to redesign an adaptive switching controller to help the system to achieve stability.
	
	According to the super twisting algorithm, the control law can be written as
	
	\begin{equation}
		\label{eqn_17}
		u_{\mathrm{HOSMC}}=u_{1}+u_{2}=-\lambda|\sigma|^{\rho} \operatorname{sign}(\sigma)-\int W \operatorname{sign}(\sigma)
	\end{equation}
	
	Then the control equation is
	
	\begin{equation}
		\label{eqn_18}
		u_{\mathrm{HOSMC}}=\left\{\begin{array}{l}{-\lambda|\sigma|^{\rho}-\int W \mathrm{d} t, \sigma>0} \\ {\lambda|\sigma|^{\rho}+\int W \mathrm{d} t, \sigma<0}\end{array}\right.
	\end{equation}
	
	Let $\hat{\boldsymbol{f}}_{r}=\hat{\mathbf{M}}_{e}(\boldsymbol{e}) \ddot{\boldsymbol{e}}_{r}+\hat{\boldsymbol{h}}(\boldsymbol{q}, \boldsymbol{e})$ be the predictable dynamic vector $\hat{\boldsymbol{f}}$'s equivalent control quantity, then the system input control quantity obtained based on the equivalent control law is
	
	\begin{equation}
		\label{eqn_19}
		u_{1}=\mathbf{J}^{\mathrm{T}}(\boldsymbol{e}) \boldsymbol{\tau}_{e}=\left\{\begin{array}{ll}{-\lambda\left|\dot{\boldsymbol{e}}-\dot{\boldsymbol{e}}_{r}\right|^{\rho}+\mathbf{J}^{\mathrm{T}}(\boldsymbol{e}) \hat{\boldsymbol{f}}_{r},} & {\sigma>0} \\ {\lambda\left|\dot{\boldsymbol{e}}-\dot{\boldsymbol{e}}_{r}\right|^{\rho}+\mathbf{J}^{\mathrm{T}}(\boldsymbol{e}) \hat{\boldsymbol{f}}_{r},} & {\sigma<0}\end{array}\right.
	\end{equation}
	
	For the system dynamics model without external disturbance, the equivalent control law of $u_1$ could meet the requirements. Otherwise, the switching control $u_2$ needs to be added adaptive characteristic in order to achieve robustness.
	
	\subsection{The adaptive control law}
	
\begin{figure}[htb]
		\centering
		\includegraphics[width=0.9\textwidth]{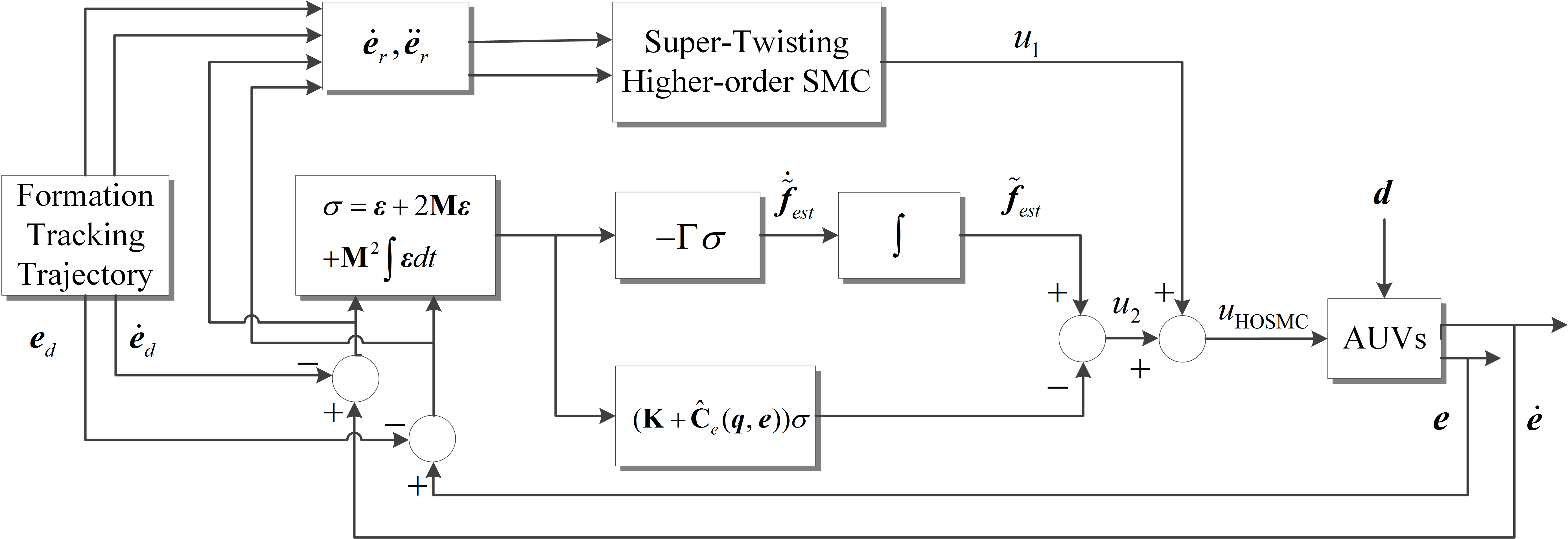}%
		\caption{Block diagram of adaptive higher-order sliding mode controller.}
		\label{fig_5}
	\end{figure}

	By adding an adaptive control link to the higher-order sliding mode controller, we can obtain a chattering-free and anti-interference controller. The controller's block diagram is shown in Fig. \ref{fig_5}.
	
	As mentioned above, the switching control $u_2$ becomes the adaptive controller after adding the adaptive characteristic to it. In order to replace the discontinuous function in the switching controller, we use a continuous adaptive control law
	
	\begin{equation}
		\label{eqn_20}
		u_{2}=\mathbf{J}^{\mathrm{T}}(\boldsymbol{e})\left(\tilde{\boldsymbol{f}}_{e s t}-\left(\mathbf{K}+\hat{\mathbf{C}}_{e}(\boldsymbol{q}, \boldsymbol{e})\right) \sigma\right)
	\end{equation}
	where $\mathbf{K} \in \mathbb{R}^{6 \times 6}$ is a positive definite diagonal matrix relative to the convergence rate of the controller. $\tilde{\boldsymbol{f}}_{e s t}$ is an adaptive term realizing the estimation of lumped parameter uncertainty vector $\tilde{\boldsymbol{f}}$ in (\ref{eqn_6}). The renewal rate of $\tilde{\boldsymbol{f}}$ is designed as
	
	\begin{equation}
		\label{eqn_21}
		\dot{\tilde{\boldsymbol{f}}}_{e s t}=-\boldsymbol{\Gamma} \sigma
	\end{equation}
	where $\boldsymbol{\Gamma} \in \mathbb{R}^{6 \times 6}$ is the positive definite diagonal matrix related to the adaptive rate. The adaptive term $\tilde{\boldsymbol{f}}_{est}$ is the error estimation of sliding mode surface $\sigma$, which can make the predicted system dynamic state more close to the actual system dynamic state under unknown disturbances. If $\tilde{\boldsymbol{f}}_{est}$ is bounded, $\dot{\tilde{\boldsymbol{f}}}_{est}$ is also bounded.
	
	\begin{myAssumption}
		Let $\boldsymbol{w}=\tilde{\boldsymbol{f}}_{e s t}-\tilde{\boldsymbol{f}}$. If and only if $\dot{\tilde{\boldsymbol{f}}}^{\mathrm{T}} \boldsymbol{\Gamma}^{-1} \boldsymbol{w}<0$, the following inequality is true:
		\begin{equation}
			\label{eqn_22}
			\sigma^{\mathrm{T}}\left(\tilde{\mathbf{M}}_{e}(\boldsymbol{e})+\mathbf{K}\right) \sigma \geq\left|\dot{\tilde{\boldsymbol{f}}}^{\mathrm{T}} \boldsymbol{\Gamma}^{-1} \boldsymbol{w}\right|
		\end{equation}
		\label{assumption_1}
	\end{myAssumption}
	
	\begin{myTheorem}
		For the nonlinear dynamic system described in (\ref{eqn_3}), assuming that the uncertain lumped parameter vector $\tilde{\boldsymbol{f}}$ is bounded and satisfies Assumption. \ref{assumption_1}, the whole closed-loop system is asymptotically stable after using the adaptive controller in Fig. \ref{fig_5}.
		\label{theory_1}
	\end{myTheorem}
	
	\begin{myProof}
		See Appendix A.
	\end{myProof}
	
	If the speed of the formation is slow and $\tilde{\boldsymbol{f}}$ is relatively small, then $\boldsymbol{w}$ decreases with the effect of adaptive control law. Then the inequality in Assumption. \ref{assumption_1}  is easier to implement. Even in the worst case, condition in (\ref{eqn_22}) can be meet by adjusting $\mathbf{K}$ and $\mathbf{\Gamma}$.

	\subsection{Dynamic model based predictive control}
	
	As we can model the AUV's dynamics in the physical engine simulation system, a kind of model predictive control method is combined into the adaptive higer-order sliding mode controller to make the control output more smoother.
	
	\subsubsection{Propeller arrangement and dynamic modeling}
	
	For an AUV, the control forces are usually generated by the thrusters. Different thruster allocations result different control outputs and need different control input signals \cite{gan_qpso-model_2018}. $\boldsymbol{\tau} \in \mathbb{R}^{6 \times 1}$ is the control forces and torques generated by the actuator. All the forces and torques are measured relative to the center of mass of the AUV. Normally, an AUV can have 1 to 5 thrusters. Combined with the action of the servos, these thrusters generate five degrees of freedom forces (roll degree is not considered here). The forces and torques generated by the thruster are indicated by matrix $\boldsymbol{u}_t \in \mathbb{R}^{P_t}$, where $P_t$ is the number of thrusters. Then we have
	
	\begin{equation}
		\label{eqn_23}
		\boldsymbol{\tau}=\boldsymbol{B}_{t} \boldsymbol{u}_{t}
	\end{equation}
	where $\boldsymbol{B}_{t} \in \mathbb{R}^{6 \times {P_t}}$ is the thruster control matrix (TCM). Because the AUV studied is usually underactuated, we have $P_t < 6$.
	
\begin{figure}[tbh]
		\centering
		\subfigure[]{\includegraphics[width=0.5\textwidth]{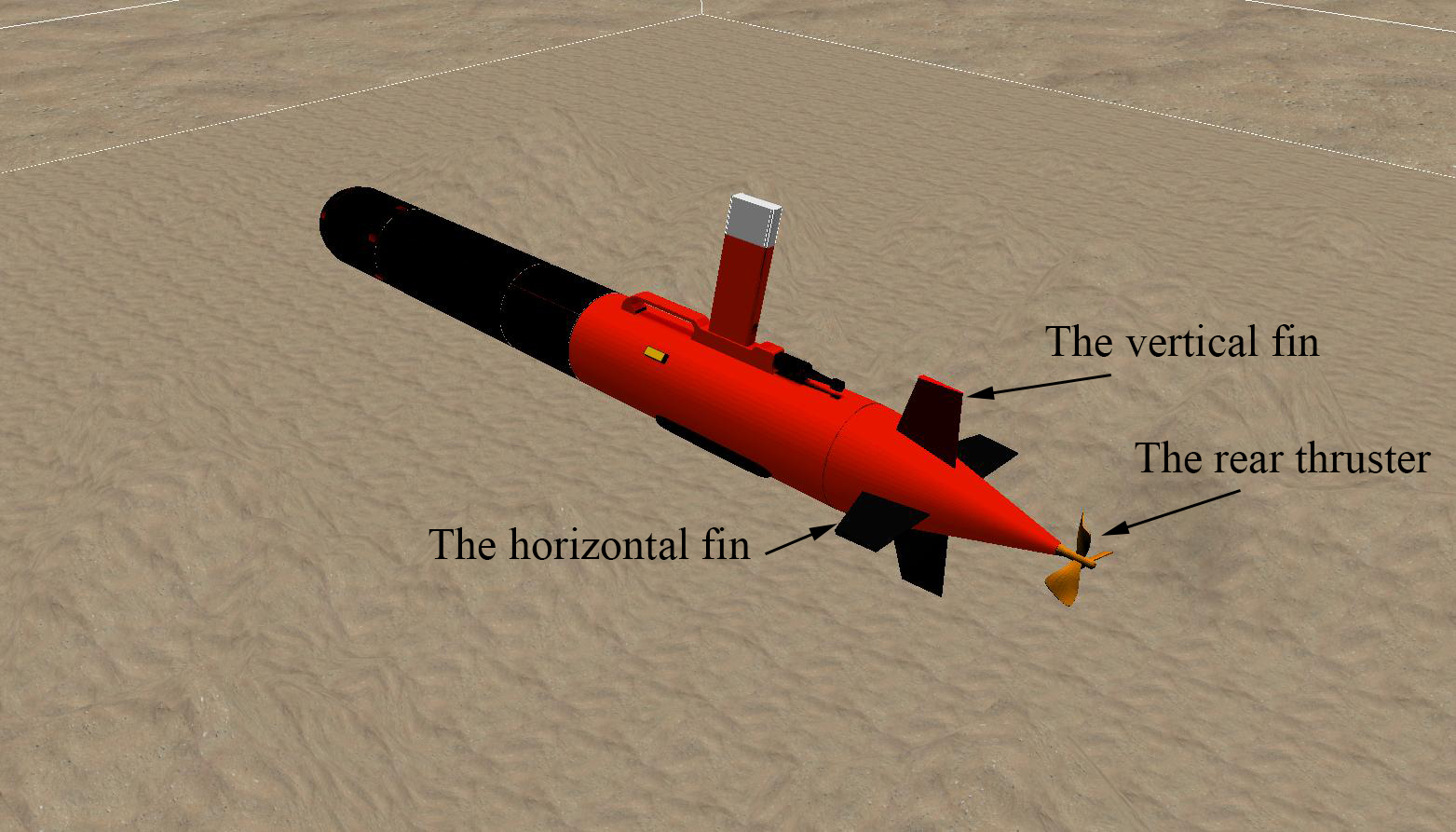} \label{fig_6a}}%
		\subfigure[]{\includegraphics[width=0.4\textwidth]{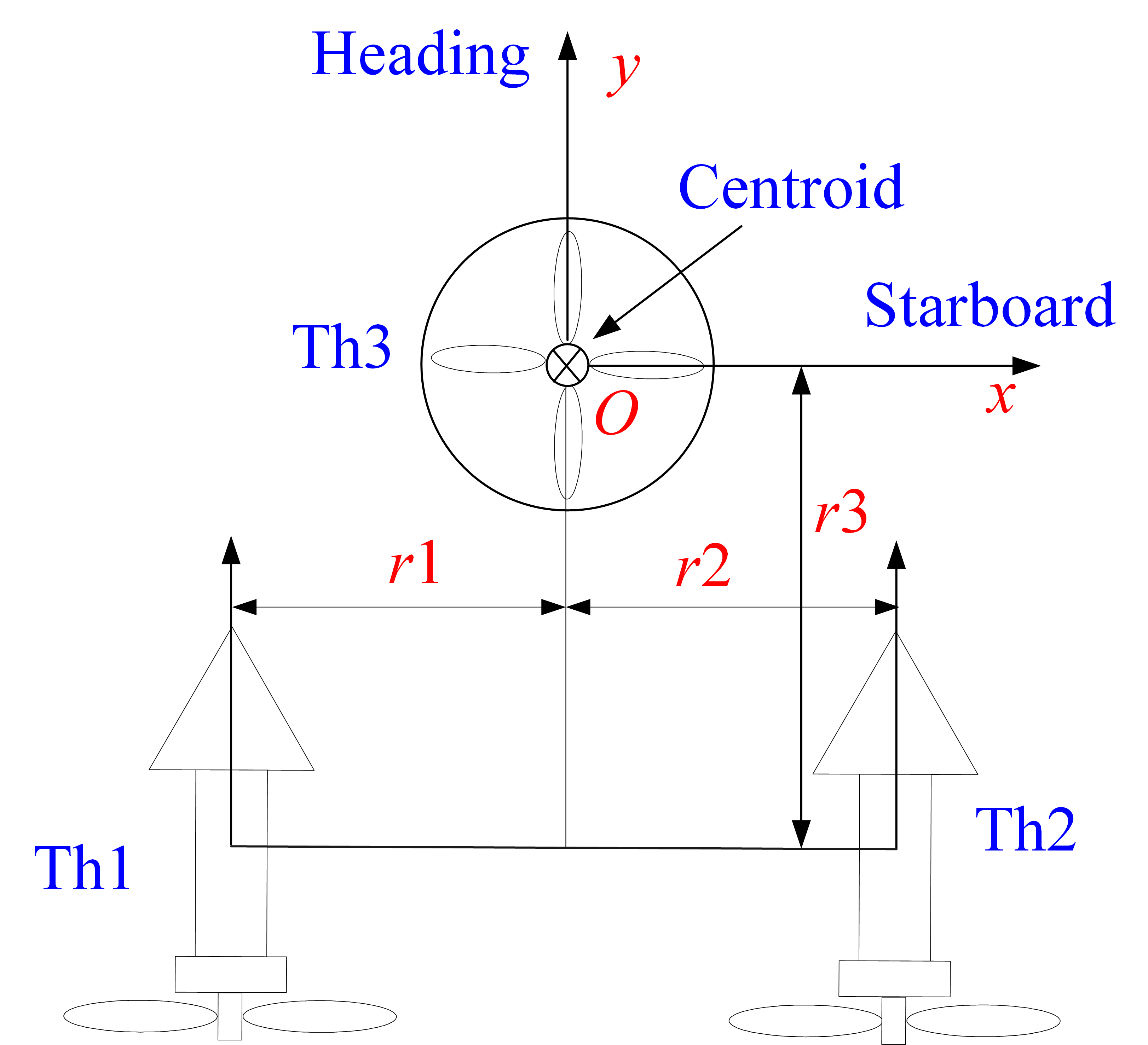} \label{fig_6b}}
		\caption{Thruster configuration for the experiment platform. (a) The LAVU's dynamic configuration, (b) the equivalent three thruster model.}
		\label{fig_6}
	\end{figure}

	The open-frame AUV named ``LAUV'' is the research object for formation control in this paper.
	This kind of AUV has one horizontal thruster and four symmetrically located fins (two vertical and two horizontal), forming an underactuated controller model, as shown in \mbox{Fig. \ref{fig_6a}} \cite{SOUSA2012268}.
	This mechanical configuration leads to a simple dynamic model. The dynamics of the thruster motor and fin servos are generally much faster than the remaining dynamics, therefore, they can be excluded from the model.
	The LAUV is also symmetric in shape. For safety reasons, it usually is slightly buoyant. The center of gravity is slightly below the center of buoyancy, providing a restoring moment in pitch and roll which is useful for these underactuated vehicles.
	Traditionally, three parameters including propeller velocity, horizontal fin inclination, and vertical fin inclination, are considered in modelling. In this work, we compare the fin' force and moment from \cite{silva_2007} with a traditional thruster' force and moment, and find that the LAUV's dynamic model could be equivalently represented with a three thruster model shown in \mbox{Fig. \ref{fig_6b}}, where TH\textit{i} (\textit{i}=1,2,3) represents the \textit{i}th thruster. In this way, the proposed sliding mode controller could be directly applied to the thrusters without considering the fin's servo control variables and transition formulas, which facilitates our physics engine simulations.
	The coupling relationship between control quantity and the propellers can be obtained according to this kind of model making the representation of forces and moments more direct. After that, we implemented simulations with the three thrust model in a way that the physical dynamic can be included. Then we could generate the control equations of the three thrusters.
	
	In the 3-D environment, the LAUV moves with 6-DOF under underactuated conditions. For the AUV operation or simulations, the rolling action could be neglected temporarily as $\tau_p = 0$. Then the dynamic parameters of the AUV can be obtained by decoupling as $\boldsymbol{\tau}=\left[\tau_{u} \ {\tau_{v}} \ {\tau_{r}} \ {\tau_{w}} \ {\tau_{q}} \right]^{\mathrm{T}}$ with $\boldsymbol{e}=\left[{x} \ {y} \ {\psi} \ {z} \ \theta \right]^{\mathrm{T}}$ accordingly. Let $K_i$ be the propellers' control parameters, and $u_{Thi}$ be the $i$\textit{th} thruster's force, $i = 1,2,3$. According to (\ref{eqn_23}) and the propeller configuration shown in \mbox{Fig. \ref{fig_6}}, The control force and torque equation can be written as (\ref{eqn_24}).
	
	\begin{equation}
		\label{eqn_24}
		\boldsymbol{\tau}=\left[\begin{array}{c}{\tau_{u}} \\ {\tau_{v}} \\ {\tau_{r}} \\ {\tau_{w}} \\ {\tau_{q}}\end{array}\right]=\boldsymbol{B}_{t} u_{t}=
		\left[\begin{array}{ccc}{K_{1} L_{1}} & {K_{2} L_{2}} & {0} \\
			{-t_{1}\left(1-K_{1}\right) L_{1}} & {t_{2}\left(1-K_{2}\right) L_{2}} & {0} \\
			{K_{1} L_{1} r 1} & {-K_{2} L_{2} r 2} & {0} \\
			{0} & {0} & {K_{3}} \\
			{t_{3} K_{1}\left(1-L_{1}\right) r 3} & {t_{4} K_{2}\left(1-L_{2}\right) r 3} & {0}\end{array}\right]
		\left[\begin{array}{c}{u_{T h 1}} \\ {u_{T h 2}} \\ {u_{T h 3}}\end{array}\right]
	\end{equation}
	
	In (\ref{eqn_24}), let $K_{1}, K_{2} \in[0.2 \ 1]$, $K_{3} \in [-1 \ 1]$, $L_{1}, L_{2} \in[0.3 \ 1]$, $t_{1}, t_{2} \in[0 \ 1]$, and $t_{3}, t_{4} \in[-0.5 \ 0.5]$, which are all adjustable coefficients. At the same time, in \mbox{(\ref{eqn_12})} and \mbox{(\ref{eqn_13})}, parameters are set as that $\rho=0.36$, $W = 0.3$, $\sigma_{0}=0.1$. These parameters can be adjusted according to the situation to obtain better control results. As for the parameters of adaptive control, we set $\mathbf{K}=\operatorname{diag}(50 \ 50 \ 0 \ 0\ 0\ 50)$ and $\boldsymbol{\Gamma}=\operatorname{diag}(50 \ 50\ 0\ 0\ 0\ 100)$.
	
	\subsubsection{\textcolor{red}{Predictive control strategy combined with the SMC controller}}
	
	Model predictive control (MPC) is a multivariable control algorithm that uses an internal dynamic model of the process, a cost function over the receding horizon, and an optimization algorithm minimizing the cost function using the control input.

	In this formation control problem, we adopt MPC as a shell for the proposed SMC controller. As the SMC controller could handle the finite water flow and other disturbance, we don't worry about the single propeller's control input and output. Howerver, sudden changes on large temporal and spatial scales are intractable by the SMC controller in some complex water flow environments. MPC can be used to some extent for preprocessing of sudden and dramatic changes in perturbation. In this way, the SMC controller can avoid abrupt input changes on the sliding surface.
	
\begin{figure}[!h]
		\centering
		\includegraphics[width=0.95\textwidth]{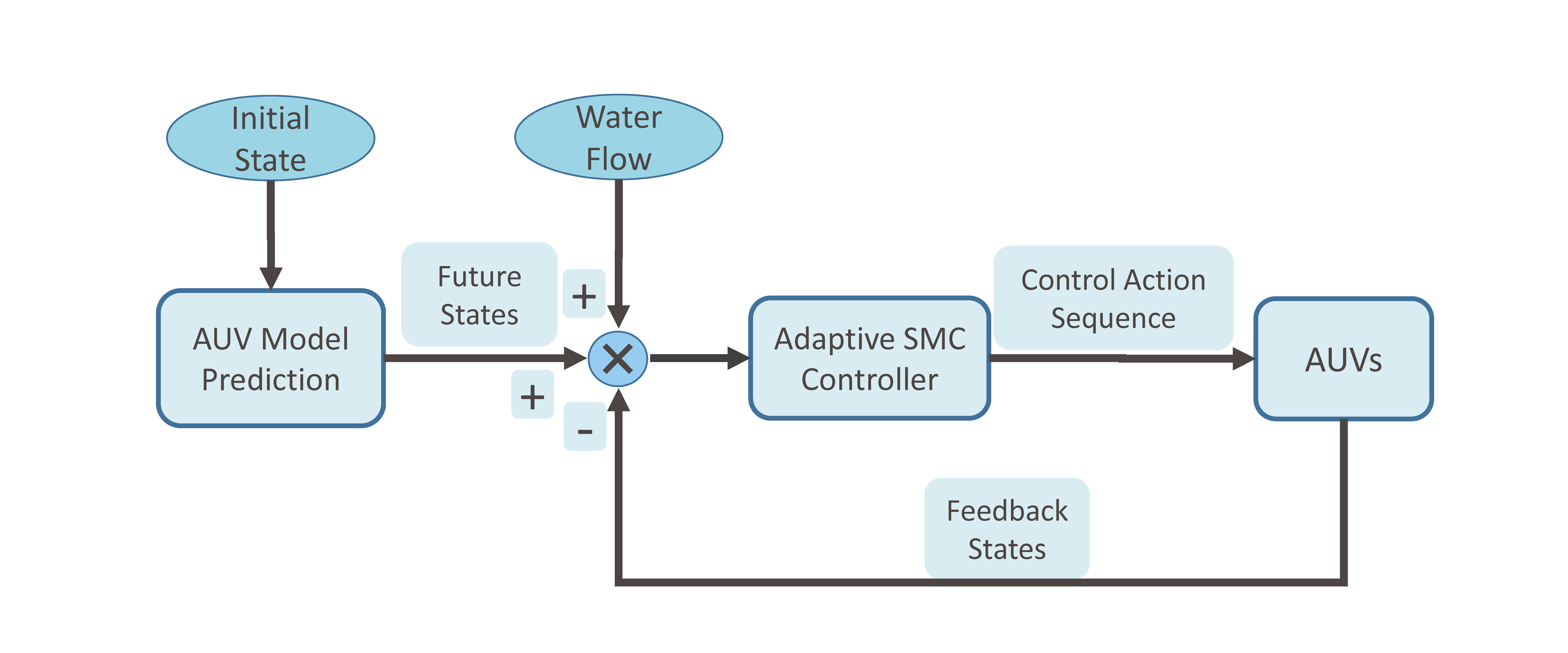}%
		\caption{Diagram of the model predictive control flow.}
		\label{fig_mpc}
	\end{figure}

	The MPC control flow mainly consists of two main blocks. As shown in \mbox{Fig.\ref{fig_mpc}}, one is the prediction of future system behavior on the basis of current measurements and a system model (hence referred to as AUV model prediciton and feedback states), and the other is solution of an optimization for determining future values of the manipulated variables, subject to constraints (hence referred to as control action sequence).
	
	In the AUV's model prediction, the future system output in the prediction domain is given as \\
	\begin{equation}
		\label{eqn_mpc_1}
		\min J=\sum_{k=1}^{N_{e}}\left\|\dot{\boldsymbol{e}}(t)-\dot{\boldsymbol{e}}_{d}(t+k)\right\|^{2}+\sum_{k=1}^{N_{e}}\sum_{i=1}^{N_{u}}\|\boldsymbol{u}(t+k)-\boldsymbol{u}(t+k+i)\|^{2}
	\end{equation}
	\begin{equation*}
		\begin{aligned}
			\text { s.t. } & \dot{\boldsymbol{e}}(t+1)=f(\dot{\boldsymbol{e}}(k), \boldsymbol{\tau}(t+k))  & (\mathrm{a})\\
			& \boldsymbol{\tau}(t) \in[A, \bar{A}], t \in T & (\mathrm{b})\\
			& \boldsymbol{e}(t) \in[B, \bar{B}], t \in T)  & (\mathrm{c})
		\end{aligned}
	\end{equation*}
	
	In \mbox{(\ref{eqn_mpc_1})}, $\dot{\boldsymbol{e}}(t)$ is quoted as the tracking error's first order derivative obtained by the solution of system state at the current moment. $\dot{\boldsymbol{e}}_{d}(t+k)$ is expressed as the expected tracking error's first order derivative as the system state at the $k$th prediction time, where $k \in [1,N_{e}]$. $\boldsymbol{u}(t+k)$ is the controller's action related to $\boldsymbol{\tau}$ in \mbox{(\ref{eqn_24})} at the corresponding moment of the $k$th prediction. $\boldsymbol{u}(t+k+i)$ is the $i$th control matrix of the SMC controller at a instant predction time, where $i \in [1,N_{u}]$. In \mbox{(\ref{eqn_mpc_1})}, constraint (a) represents the dynamic characteristics of the controlled object which related to $\boldsymbol{\tau}(t+k)$; (b) and (c) represent the control quantity and the state quantity, respectively subject to an upper and lower bound. In real world simulation, the upper and lower bound mainly means the maximum force and torque of the thruster, and the maximum acceptable errors for formation tracking.
	
	\section{Simulation Results}
	
	The simulation study is based on the algorithm proposed above. In this study, we sample and simplify the water flow model data as provided in section III, which hourly reflect the current situations in the real world. Physics simulation engine based experiments in Gazebo are adopted as well as in the MATLAB software. The UUV Simulator \cite{7761080} is employed mainly for the 3-D physical simulations of the  AUVs' model based on the ``LAUV" vehicle.
	The simulation scenario is designed to explain how the algorithm is working. The water flow speed and direction is designed alterable but the absolute value of speed is limited to a range of $[0 \ 0.5]$ ($m/s$), which is slower than the AUVs' speed, but changing in directions. The area is limited to a cubic water flow field of the 3-D workspace [$(x,y) \in [0 \ 80]$ ($m$), and $(z) \in [-20 \ 0]$ ($m$)].
	\subsection{Formation tracking results}
	
	Considering a team of 3 AUVs in a triangular formation tracking an arbitrary curve path, we assume that the leader AUV moves first and all the followers have essential sensors to contact with the leader and acquire essential informations such as relative distances and angles to the leader. The control object is to make the followers track the desired paths simultaneously as accurate as possible.
	The whole formation is kept and the path is tracked in this way.
	The controller adjusts the velocities and the directions of the AUVs in the formation to achieve this goal.
	In the simulations, we set the time-varied water flow disturbance vector ${\boldsymbol{d}_o}=[{C_x} \ {C_y} \ {0} \ {0} \ {0} \ {C_z}]$ equivalent to the parameter $\boldsymbol{d}$ depicted in \mbox{Fig. \ref{fig_5}}. In this way the water flow influence is resolved into three quantities in the $x,y,z$ directions respectively, which is convenient for simulation.
	As the current velocity is assumed to be bounded, let $\left(C_{x}, C_{y}, C_{z}\right) \in[-20 \ 20]$ $(\mathrm{N})$.
	
	Based on the formation control algorithm, 3 AUVs form a triangular formation at the starting position. The leader runs along the desired path and meanwhile generates the desired path of the 2 followers. The follower AUV also runs along the desired path according to the adaptive higher-order sliding mode control algorithm, until the formation reaches the target position. During the tracking procedure, the trajectory tracking and formation keeping are both well maintained. As shown in \mbox{Fig. \ref{fig_7}}, the formation moves along the spiral trajectory in triangular shape. The formation tracking test achieves good results.
	
	\begin{figure}[htb]
		\centering
			\subfigure[]{\includegraphics[width=0.7\textwidth]{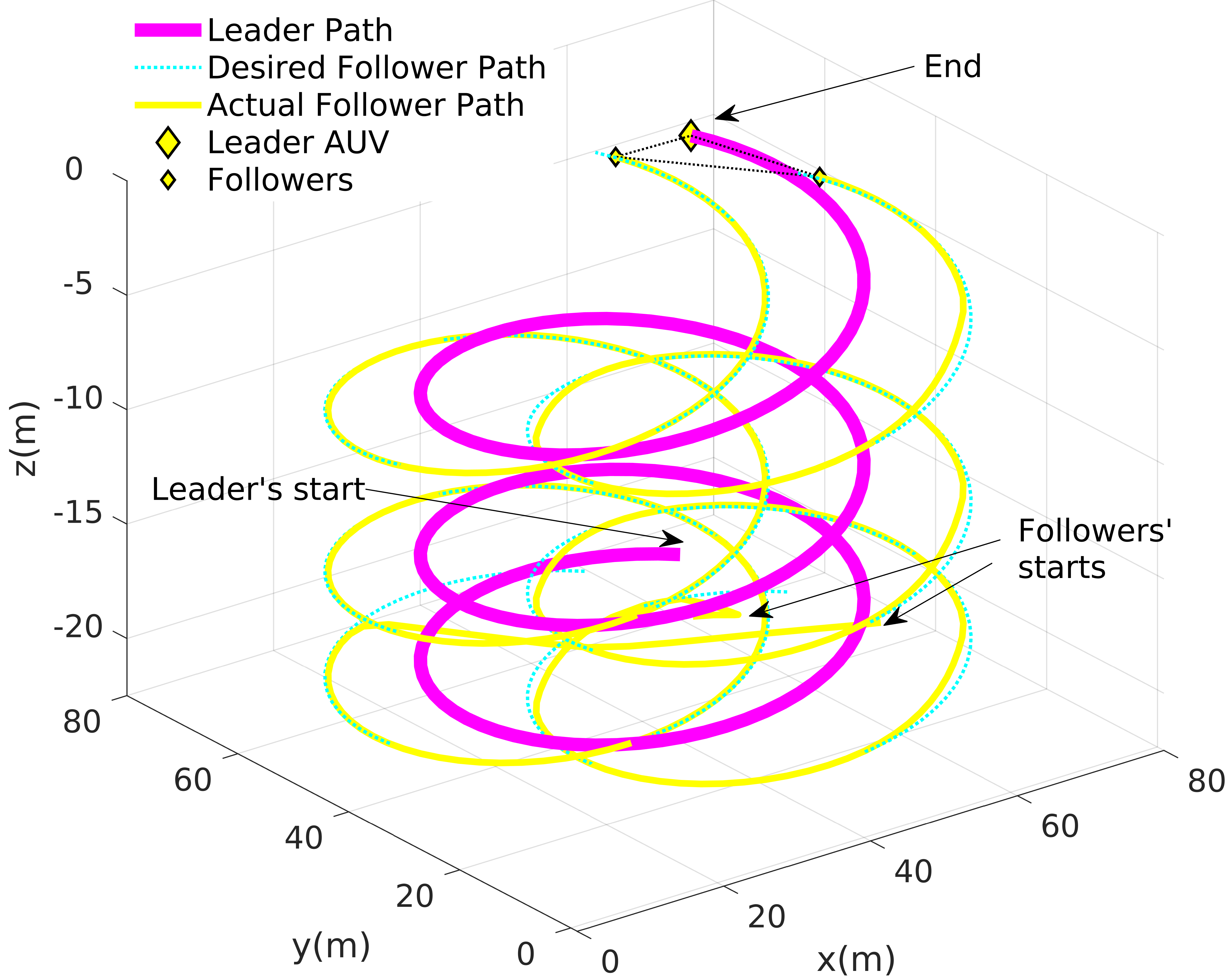} \label{fig_7}}%
		\subfigure[]{\includegraphics[width=0.45\textwidth]{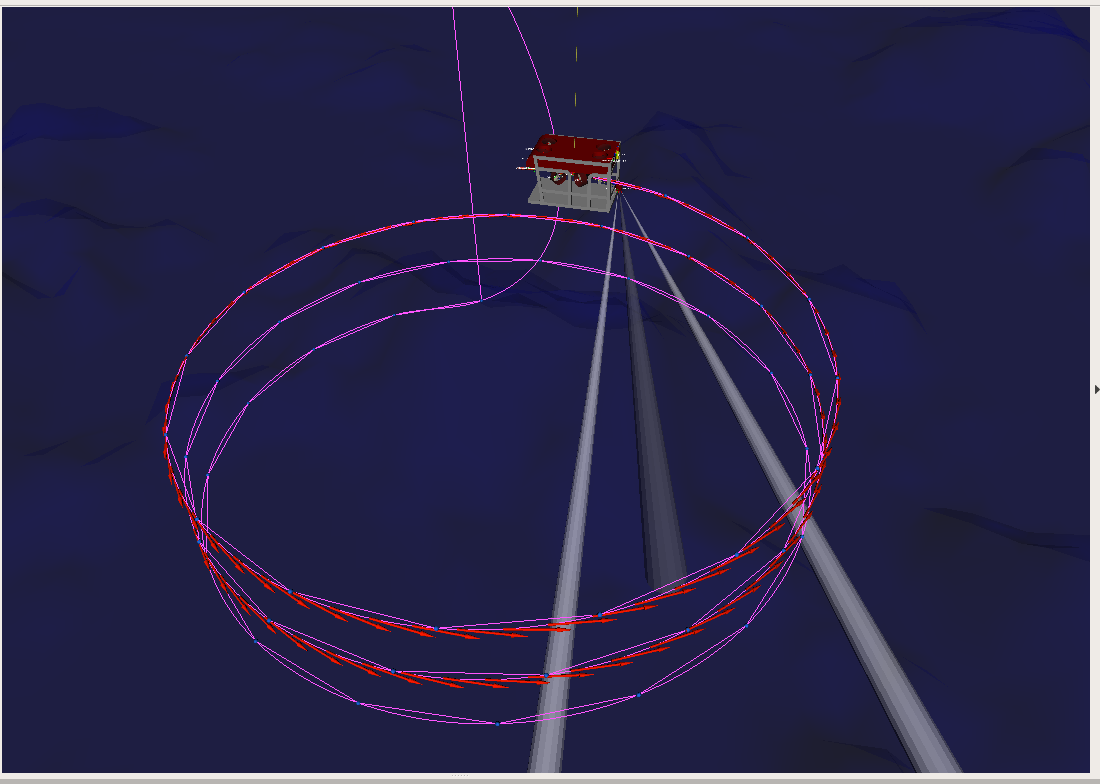} \label{fig_7b}}
		\caption{Simulation results of AUV formation tracking under 3-D water flow environment, (a) based on MATLAB, (b) based on physical engine of Gazebo 7 with ROS}
		\label{fig_7}
	\end{figure}
	
\begin{figure}[!htb]
		\centering
		\includegraphics[width=0.6\textwidth]{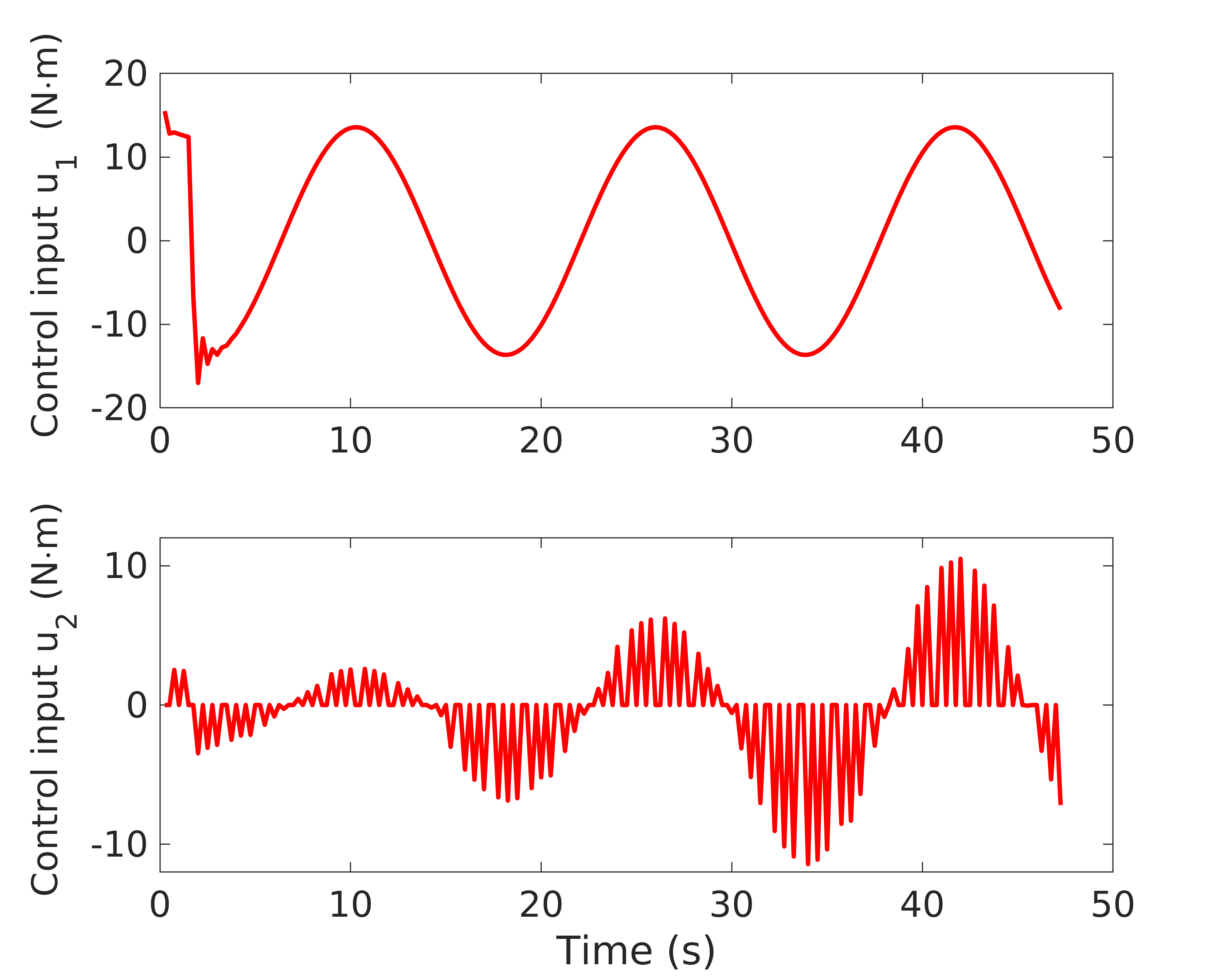}%
		\caption{$x$-axis controller input in 3-D environment.}
		\label{fig_8}
	\end{figure}

	Taking the left follower as the research object, the control effects in three directions of $x$, $y$ and $z$ are analyzed, i.e. $\tau_u$, $\tau_v$ and $\tau_w$ correspondingly. $\tau_u$ comprises two parts of $u_1$ and $u_2$ as mentioned above. The input of the controller in $x$ direction is show in \mbox{Fig. \ref{fig_8}}.
	
\begin{figure}[!htb]
		\centering
		\includegraphics[width=0.6\textwidth]{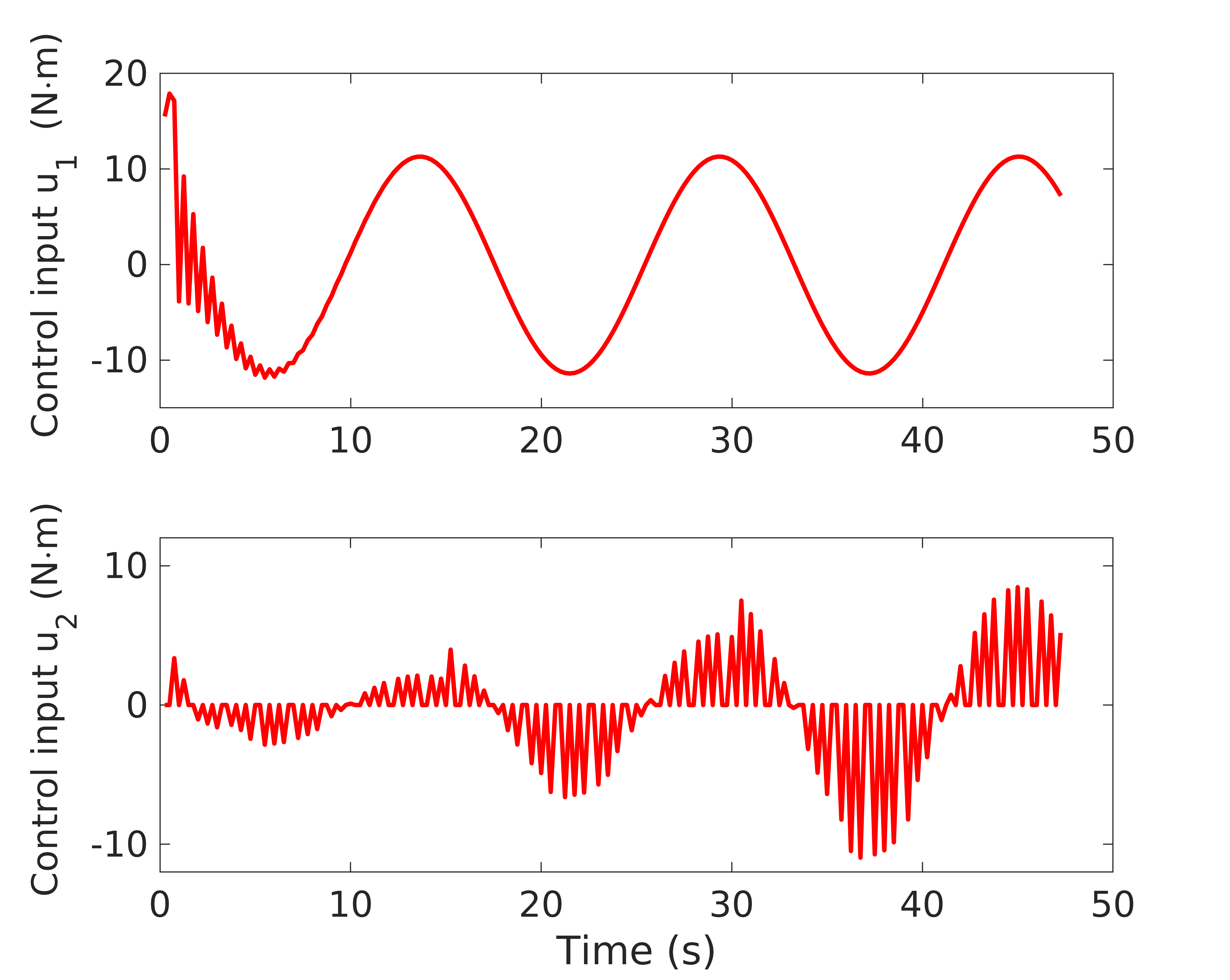}%
		\caption{$y$-axis controller input in 3-D environment.}
		\label{fig_9}
	\end{figure}

\begin{figure}[htb]
		\centering
		\includegraphics[width=0.6\textwidth]{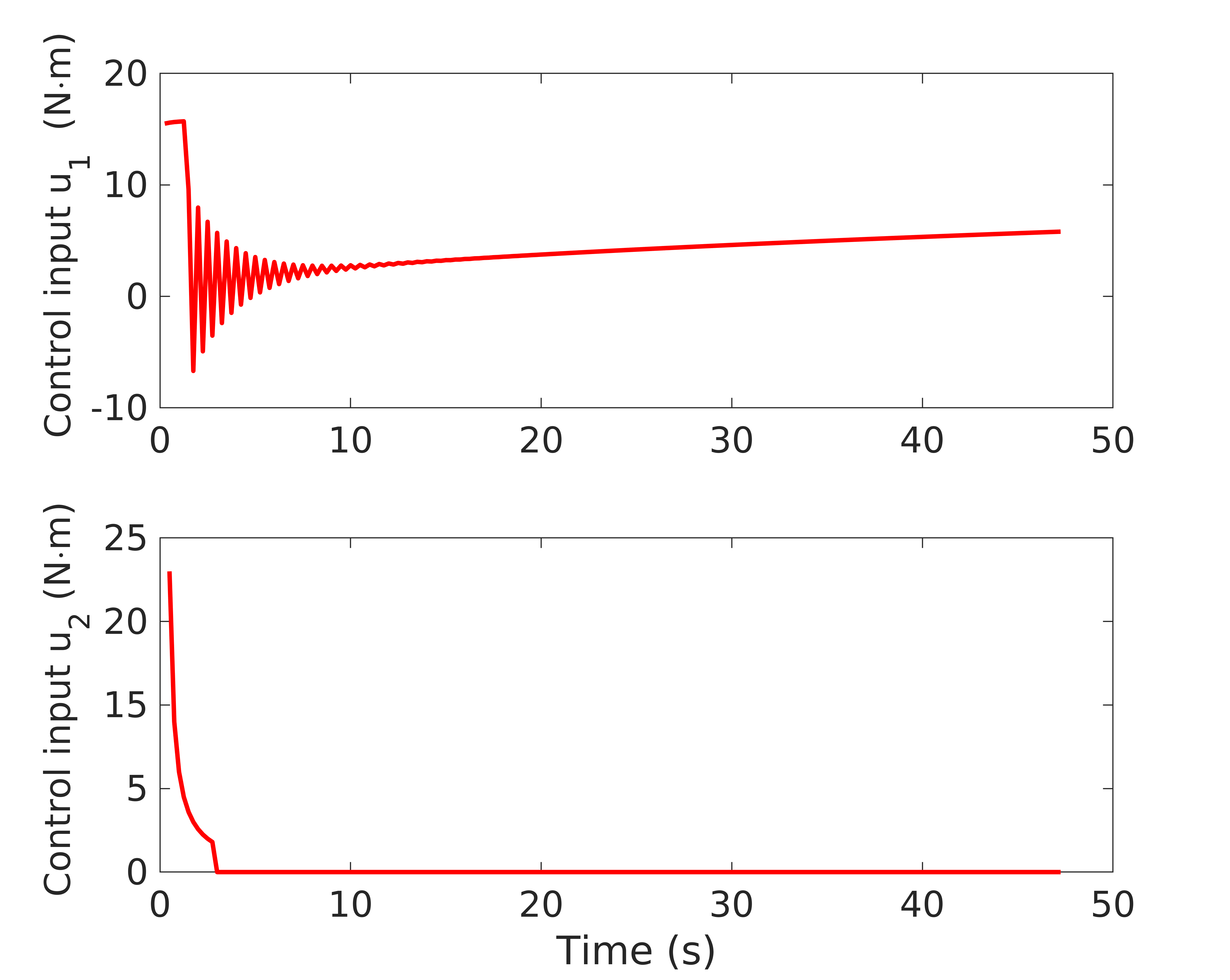}%
		\caption{$z$-axis controller input in 3-D environment.}
		\label{fig_10}
	\end{figure}

	Similarly, the controller input in the $y$ direction is also comprised of $u_1$ and $u_2$ as shown in \mbox{Figure. \ref{fig_9}}.
	In the $z$ direction, the controller input is shown in \mbox{Figure. \ref{fig_10}}.
	From the simulation results, we could find that the contoller input $u_1$ are bounded and smooth, and $u_2$ are bound pulses which are realizable.
	Unlike the control inputs in the $x$ and $y$ directions, $u_2$ in $z$ direction tends to be zero.
	This is because the water flow is layered in the 3-D space. The forces of the flow is parallel with the $x-y$ plane without vertical influence in the $z$ axis.
	
	\begin{figure}[htb]
		\centering
		\subfigure[]{\includegraphics[width=0.45\textwidth]{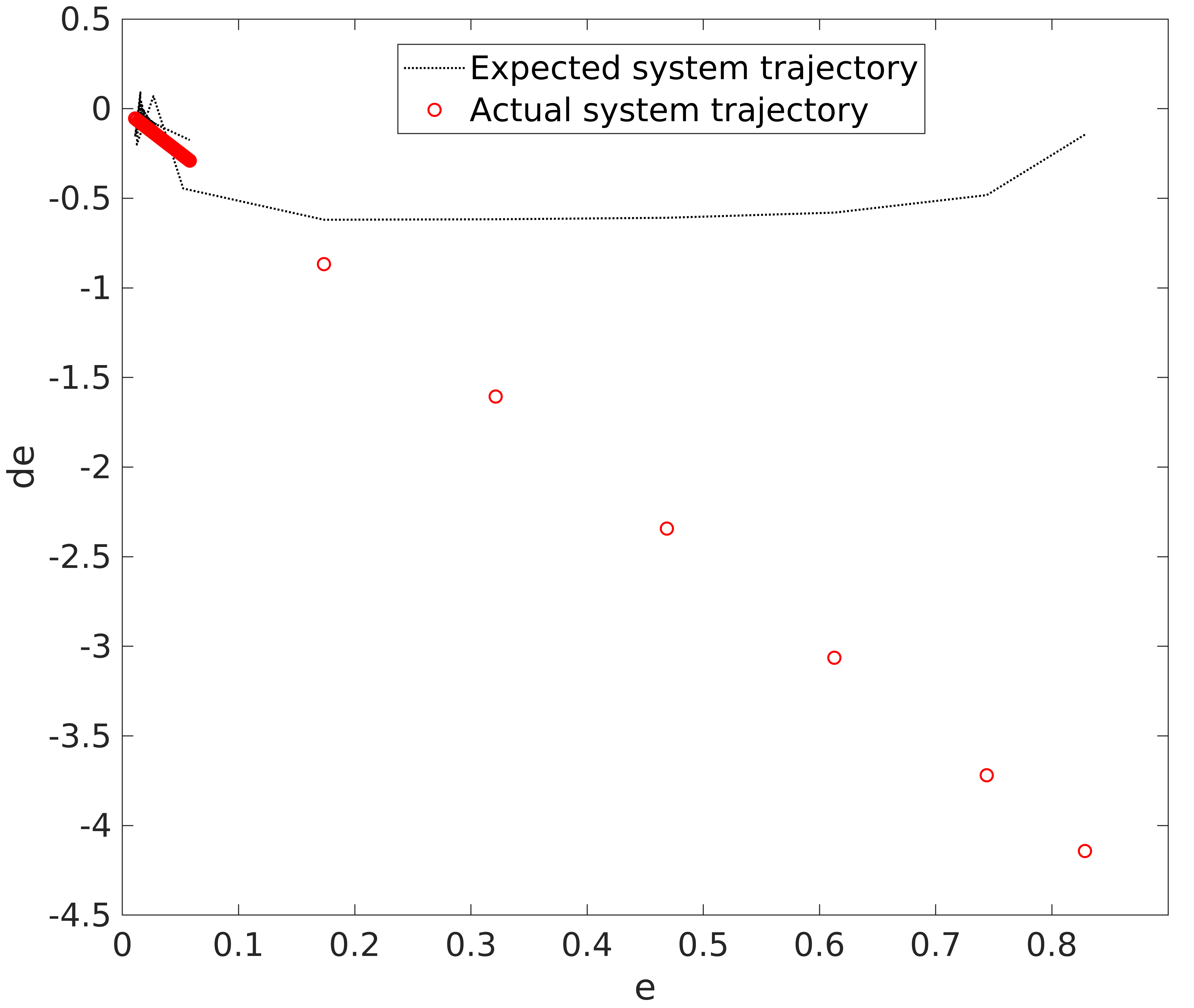} \label{fig_11a}}
		\subfigure[]{\includegraphics[width=0.45\textwidth]{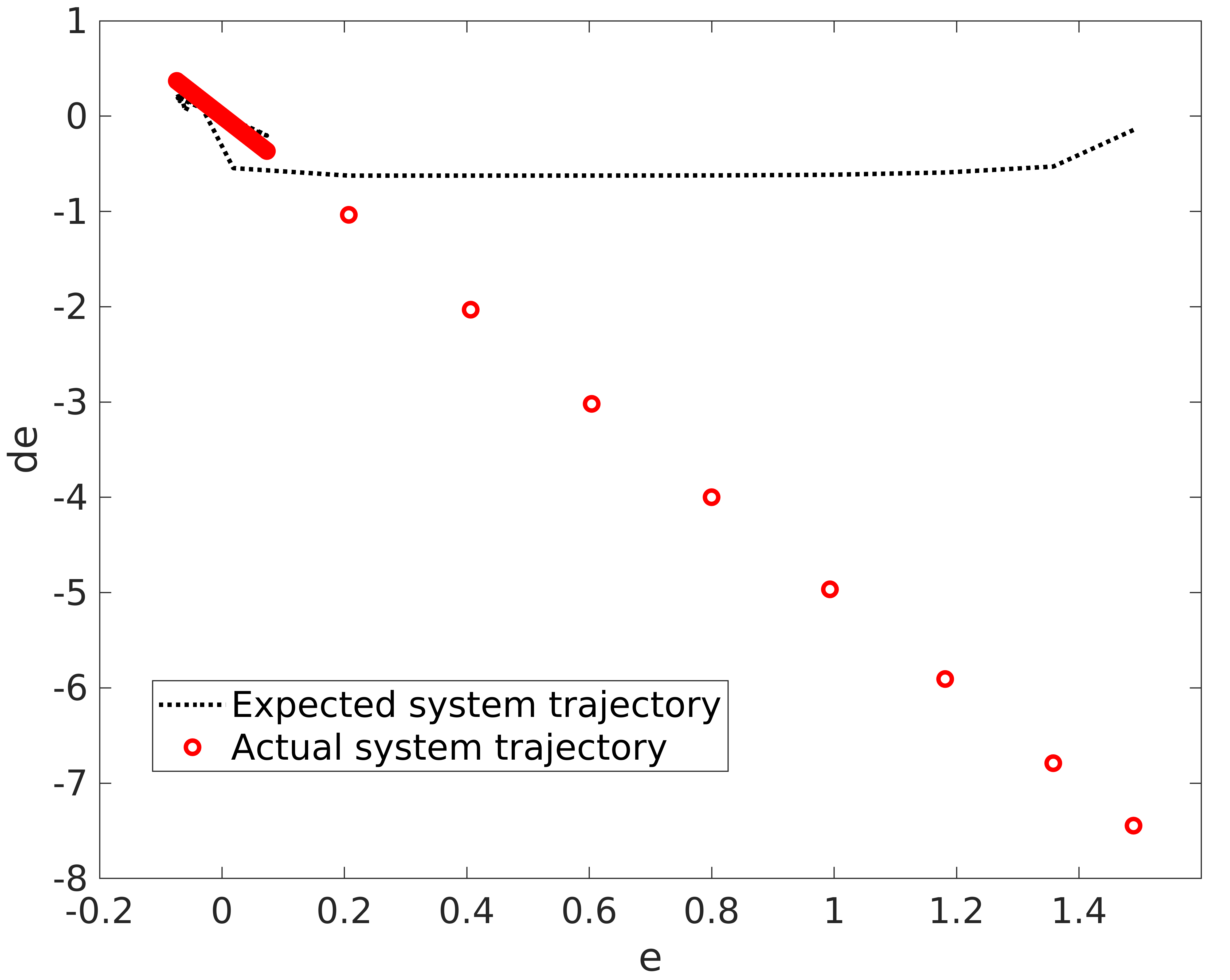} \label{fig_11b}}
		\subfigure[]{\includegraphics[width=0.45\textwidth]{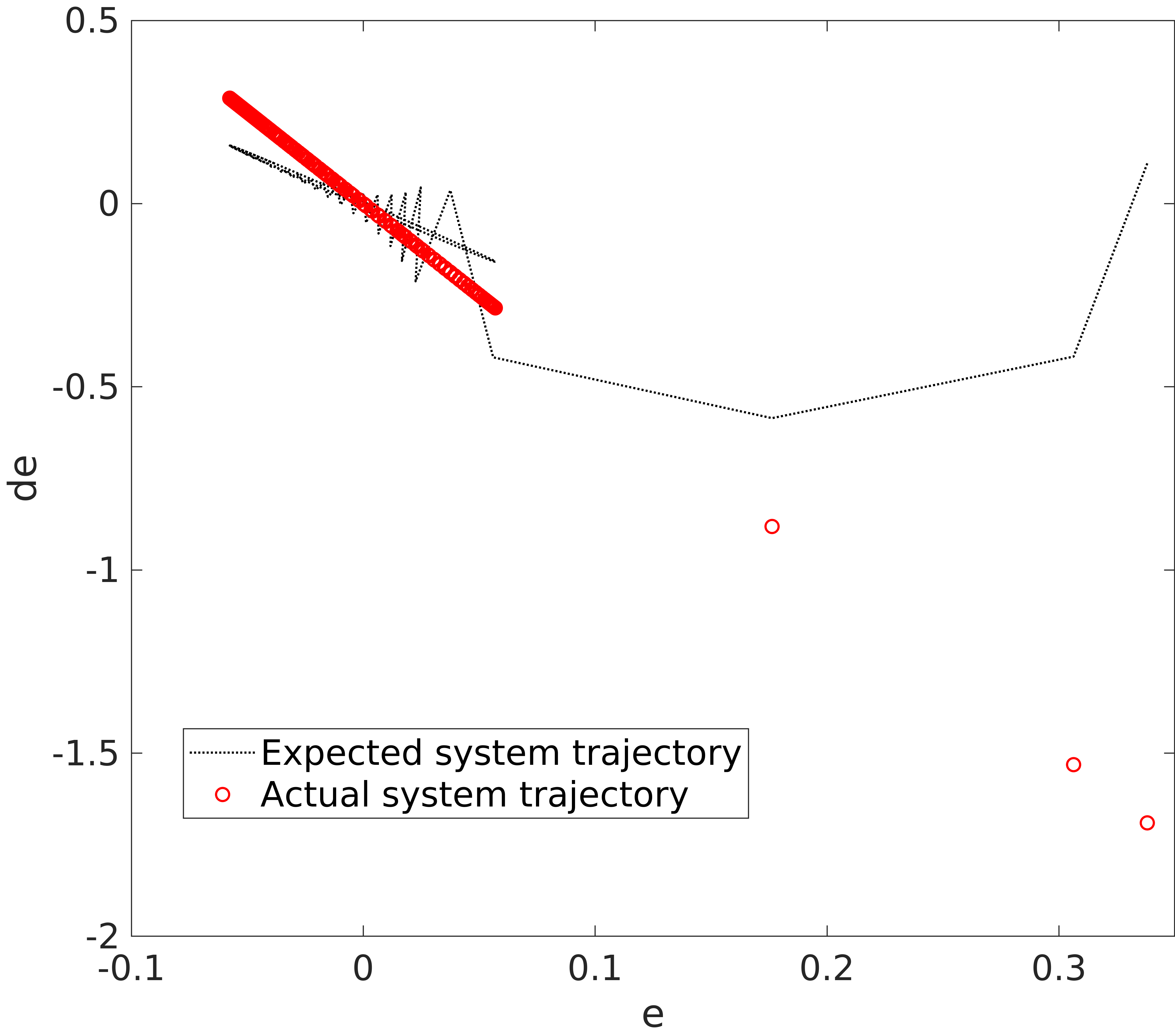} \label{fig_11c}}
		\caption{System phase trajectories of the controller in 3-D environment. (a) $x$ direction, (b) $y$ direction, (c) $z$ direction.}
		\label{fig_11}
	\end{figure}
	
	%The author names and affiliations could be formatted in two ways:
	%\begin{enumerate}[(1)]
	%\item Group the authors per affiliation.
	%\item Use footnotes to indicate the affiliations.
	%\end{enumerate}
	%See the front matter of this document for examples. You are recommended to conform your choice to the journal you are submitting to.
	
	\subsection{Controller phase trajectories and error analysis}
	
	Helical trajectory formation tracking results are smooth and interference resistible as illustrated in Fig. \ref{fig_7}. In the formation trajectory tracking process, we use the position error value $e$ as the $x$ axis and the speed error value $de$ as the $y$ axis. Then the system phase trajectories with the proposed adaptive higher-order sliding mode controller are depicted in Fig. \ref{fig_11} in the $x$, $y$ and $z$ axis respectively.
	
	\begin{table}[!htb]
		\caption{Analysis of formation tracking errors of spiral trajectory in 3-D environment}
		\centering
		\label{table_2}
		\begin{tabular}{c c c c c c c}
			\hline
			\multirow{2}*{Axis} & \multicolumn{3}{l}{Speed error (after convergence)} & \multicolumn{3}{l}{Position error (after convergence)}\\
			\cmidrule(lr){2-4} \cmidrule(lr){5-7} & Minimum & Maximum & RMSE  & Minimum & Maximum & RMSE \\
			\hline
			$x$ & 0 & 0.10 & 0.18 & 0 & 0.05 & 0.20 \\
			\hline
			$y$ & 0 & 0.15 & 0.13 & 0 & 0.08 & 0.05 \\
			\hline
			$z$ & 0 & 0.11 & 0.11 & 0 & 0.06 & 0.10 \\
			\hline
		\end{tabular}
	\end{table}

	As show in the figure, system phase trajectories converge to the expected location after certain times of iterations. In the $x$, $y$ and $z$ directions, the system trajectories chatter a little after each movement of the formation leader. That's because the algorithm need some iterations to calculate the control input. After the calculation is finished, the trajectories converge.
	
	% \subsubsection{Formation Tracking Error Analysis}
	For the position and velocity errors in the 3 directions of $x$, $y$ and $z$, the errors are limited to a small range after the sliding mode control algorithm converges. \mbox{Table. \ref{table_2}} gives the tracking error analysis, using the root mean squared error (RMSE) to measure the velocity error and the position error from the time $t=0$. The maximum and minimum errors of the system after convergence are recorded. In this experiment, the convergence moment is $t=11$ ($s$). The velocity error and position error do not converge to zero, but fluctuate regularly within a limited range, which is acceptable in practical applications.

\begin{figure}[!htb]
		\centering
		\subfigure[]{\includegraphics[width=0.45\textwidth]{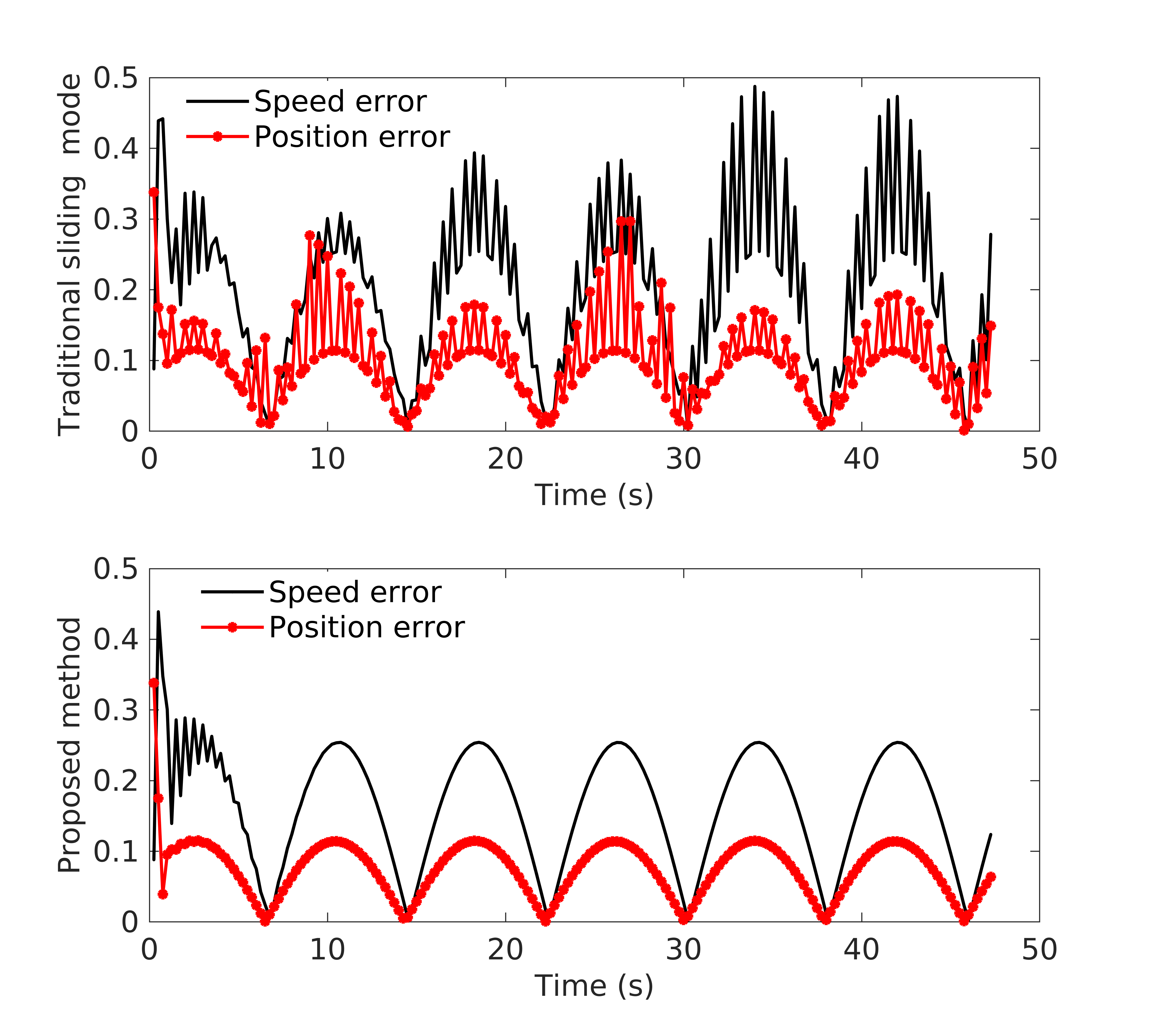} \label{fig_12a}}
		\subfigure[]{\includegraphics[width=0.45\textwidth]{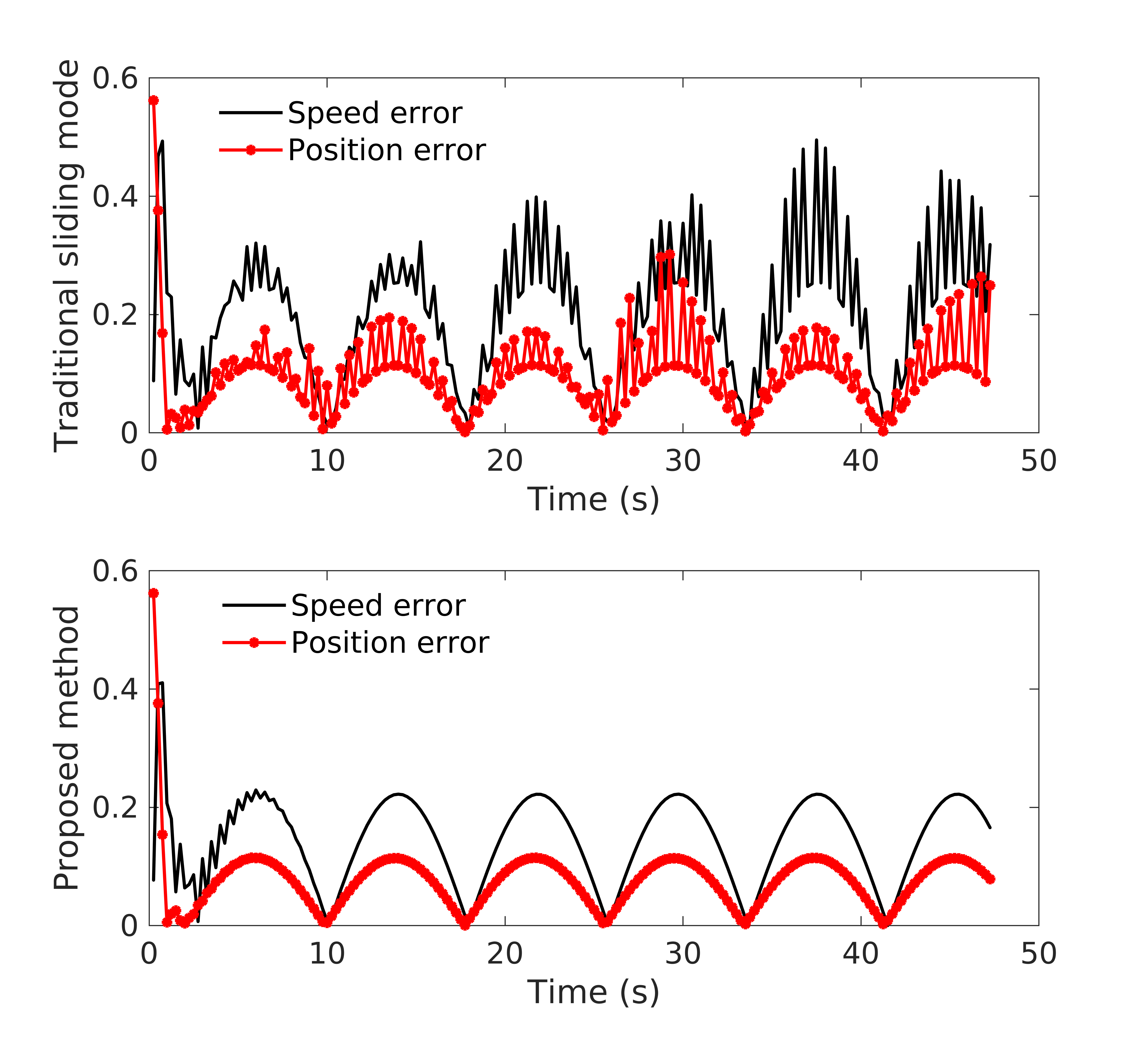} \label{fig_12b}}
		\subfigure[]{\includegraphics[width=0.45\textwidth]{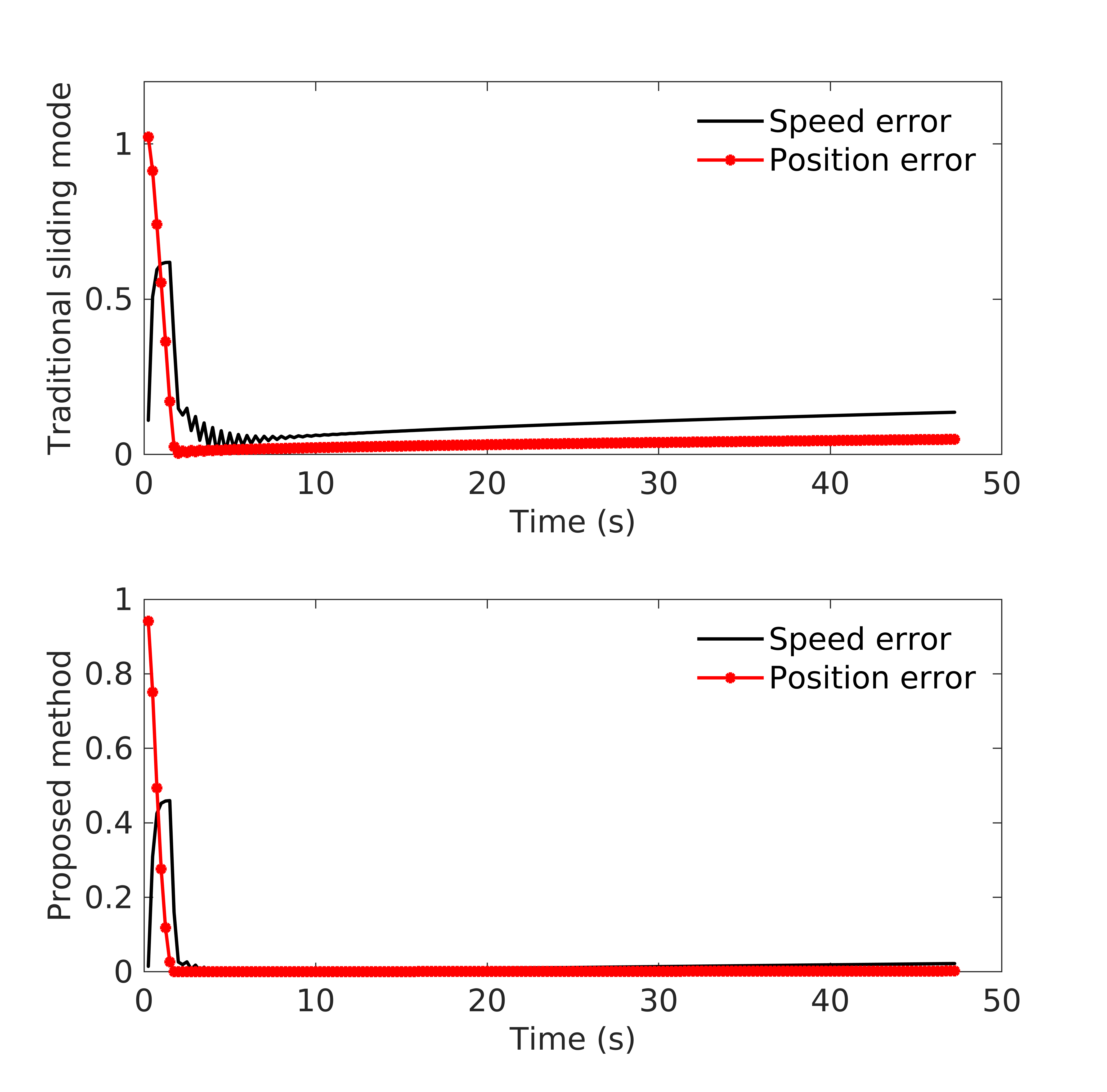} \label{fig_12c}}
		\caption{Comparison of tracking errors with the proposed method and the traditional sliding mode method. (a) $x$ direction, (b) $y$ direction, (c) $z$ direction.}
		\label{fig_12}
	\end{figure}

	\mbox{Fig. \ref{fig_12}} shows the tracking error comparison of the traditional sliding mode method and the proposed method in the paper, taking the left follower as example, too. Note that the traditional method uses the first order sliding mode controller without the adaptive module proposed in this paper. \mbox{Fig. \ref{fig_12a}} compares the tracking errors of the two kinds of methods in the $x$ axis direction. Tracking errors of the position and the speed of the AUV converge to a certain range. As the leader AUV moves, the tracking error increases and then decrease as the controller taking effect. The traditional method results jumping errors with obvious water flow influences, while the tracking error curves are smoother and smaller with the proposed method.
	In the same way, \mbox{Fig. \ref{fig_12b}} shows the tracking error comparison in the $y$ axis direction and \mbox{Fig. \ref{fig_12c}} shows the tracking error comparison in the $z$ axis direction. From these simulation results, we could find that the proposed method is effective to the formation control problem in the water flow environment.
	
	\section{Conclusion}
	
	In this paper, the formation control method based on an adaptive sliding mode control is proposed by analyzing the dynamic model of AUVs.
	A hierarchical controller structure for MAS formation is built firstly. Then an improved the higher-order sliding mode controller with the adaptive feature is designed with a model predictive controller shell on it. The controller's stability is proved by physical engine simulations.
	Based on the analysis of the water flow in 2-D and 3-D workspace, a mathematical model of the ocean current is established, and the AUV formation control simulation is carried out with flow influence added.
	The simulation results show that the AUVs in formation could maintain the relative positions in the process of trajectory tracking, and the controller has a strong ability to resist external interference.
	
	In our future work, advanced neural network method with be combined with the proposed dynamic controller to adapt it to more kinds of situations, and real-world validation of the proposed controller architecture on an existing MAS will be involved.

	

	\appendix
	\section{Proof of Theorem 1}
	
	Define a Lyapunov function as
	
	\begin{equation}
		\label{eqn_25}
		\mathrm{V}=\frac{1}{2}\left(\sigma^{\mathrm{T}} \mathbf{M}_{e}(\boldsymbol{e}) \sigma+\boldsymbol{w}^{\mathrm{T}} \boldsymbol{\Gamma}^{-1} \boldsymbol{w}\right)
	\end{equation}
	Function $\mathrm{V}$ differentiated in the time domain results
	
	\begin{equation}
		\label{eqn_26}
		\dot{\mathrm{V}}=\frac{1}{2}\left(\sigma^{\mathrm{T}} \dot{\mathbf{M}}_{e}(\boldsymbol{e}) \sigma+\dot{\boldsymbol{\sigma}}^{\mathrm{T}} \mathbf{M}_{e}(\boldsymbol{e}) \sigma+\boldsymbol{\sigma}^{\mathrm{T}} \mathbf{M}_{e}(\boldsymbol{e}) \dot{\boldsymbol{\sigma}}+\dot{\boldsymbol{w}}^{\mathrm{T}} \mathbf{\Gamma}^{-1} \boldsymbol{w}+\boldsymbol{w}^{\mathrm{T}} \boldsymbol{\Gamma}^{-1} \dot{\boldsymbol{w}}\right)
	\end{equation}
	
	According to the property of positive definite diagonal matrix, $\dot{\sigma}^{\mathrm{T}} \mathbf{M}(\boldsymbol{q}) \sigma=\sigma^{\mathrm{T}} \mathbf{M}(\boldsymbol{q}) \dot{\sigma}$, and $\dot{\boldsymbol{w}}^{\mathrm{T}} \boldsymbol{\Gamma}^{-1} \boldsymbol{w}=\boldsymbol{w}^{\mathrm{T}} \boldsymbol{\Gamma}^{-1} \dot{\boldsymbol{w}}$. From (\ref{eqn_3}), $\mathbf{M}_{e}(e) \ddot{e}=\boldsymbol{\tau}_{e}-\left(\mathbf{C}_{e}(\boldsymbol{q}, \boldsymbol{e}) \dot{\boldsymbol{e}}+\mathbf{D}_{e}(\boldsymbol{q}, \boldsymbol{e}) \dot{\boldsymbol{e}}+\mathbf{g}_{e}(\boldsymbol{e})\right)$; from (\ref{eqn_15}), $\dot{\boldsymbol{e}}=\sigma+\dot{\boldsymbol{e}}_{r}$; then we have
	
	\begin{equation}
		\label{eqn_27}
		\begin{aligned} \dot{\mathrm{V}}=
			& \frac{1}{2} \sigma^{\mathrm{T}}\left(\dot{\mathbf{M}}_{e}(e)-2 \mathbf{C}_{e}(\boldsymbol{q}, \boldsymbol{e})\right) \sigma+\\
			& \sigma^{\mathrm{T}}\left[\boldsymbol{\tau}_{e}-\left(\mathbf{M}_{e}(\boldsymbol{e}) \ddot{\boldsymbol{e}}_{r}+\mathbf{C}_{e}(\boldsymbol{q}, \boldsymbol{e}) \dot{\boldsymbol{e}}_{r}+\mathbf{D}_{e}(\boldsymbol{q}, \boldsymbol{e}) \dot{\boldsymbol{e}}+\mathbf{g}_{e}(\boldsymbol{e})\right)\right]+\dot{\boldsymbol{w}}^{\mathrm{T}} \boldsymbol{\Gamma}^{-1} \boldsymbol{w} \end{aligned}
	\end{equation}
	where $\dot{\mathbf{M}}_{e}(\boldsymbol{e})-2 \mathbf{C}_{e}(\boldsymbol{q}, \boldsymbol{e})$. It's an anti-symmetric matrix. Then $\sigma^{\mathrm{T}}\left(\dot{\mathbf{M}}_{e}(\boldsymbol{e})-2 \mathbf{C}_{e}(\boldsymbol{q}, \boldsymbol{e})\right) \sigma=0$.
	Equation (\ref{eqn_27}) can be simplified into
	
	\begin{equation}
		\label{eqn_28}
		\dot{\mathrm{V}}=\sigma^{\mathrm{T}}\left[\boldsymbol{\tau}_{e}-\boldsymbol{f}_{r}\right]+\dot{\boldsymbol{w}}^{\mathrm{T}} \boldsymbol{\Gamma}^{-1} \boldsymbol{w}
	\end{equation}
	where $\boldsymbol{f}_{r}=\mathbf{M}_{e}(\boldsymbol{e}) \ddot{\boldsymbol{e}}_{r}+\mathbf{C}_{e}(\boldsymbol{q}, \boldsymbol{e}) \dot{\boldsymbol{e}}_{r}+\mathbf{D}_{e}(\boldsymbol{q}, \boldsymbol{e}) \dot{\boldsymbol{e}}+\mathbf{g}_{e}(\boldsymbol{e})$.
	Substituting the controller input $\boldsymbol{\tau}_e$ in the inertial coordinate system with the adaptive control rate $u_2$ in the body-fixed coordinate system, and adding the predictable quantity of control $\hat{\boldsymbol{f}}_{r}$, (\ref{eqn_28}) changes to
	
	\begin{equation}
		\label{eqn_29}
		\dot{\mathrm{V}}=\sigma^{\mathrm{T}}\left(\tilde{\boldsymbol{f}}_{e s t}-(\boldsymbol{f}-\hat{\boldsymbol{f}})-\left(\mathbf{M}_{e}(\boldsymbol{e})-\hat{\mathbf{M}}_{e}(\boldsymbol{e})\right) \sigma-\mathbf{K} \sigma\right)+\dot{\boldsymbol{w}}^{\mathrm{T}} \boldsymbol{\Gamma}^{-1} \boldsymbol{w}
	\end{equation}
	
	Considering Assumption \ref{assumption_1}, (\ref{eqn_29}) reduces to
	
	\begin{equation}
		\label{eqn_30}
		\dot{\mathrm{V}}=-\sigma^{\mathrm{T}}\left(\tilde{\mathbf{M}}_{e}(\boldsymbol{e})+\mathbf{K}\right) \sigma-\dot{\tilde{\boldsymbol{f}}}^{\mathrm{T}} \boldsymbol{\Gamma}^{-1} \boldsymbol{w}
	\end{equation}
	
	By adjusting the parameters $\mathbf{K}$ and $\boldsymbol{\Gamma}$, $\left|\dot{\tilde{\boldsymbol{f}}}^{\mathrm{T}} \boldsymbol{\Gamma}^{-1} \boldsymbol{w}\right| \leq 0$, and $\sigma^{\mathrm{T}}\left(\tilde{\mathbf{M}}_{e}(\boldsymbol{e})+\mathbf{K}\right) \sigma \geq\left|\dot{\tilde{\boldsymbol{f}}}^{\mathrm{T}} \boldsymbol{\Gamma}^{-1} \boldsymbol{w}\right|$ could be both satisfied. Then $-\sigma^{\mathrm{T}}\left(\tilde{\mathbf{M}}_{e}(e)+\mathbf{K}\right) \sigma-\dot{\tilde{f}}^{\mathrm{T}} \mathbf{\Gamma}^{-1} \boldsymbol{w} \leq 0$, i.e. $\dot{\mathrm{V}} \leq 0$, which satisfies the Lyapunov stability criterion. This completes the proof of Theorem \ref{theory_1}.
	
	\vspace{2ex}
	\noindent
	{\bf\normalsize Acknowledgement}\newline
	{The authors would like to thank all the editors and reviewers who have given their valuable advice to this paper.
		This project is supported in part by the National Natural Science Foundation of China (62033009, 61873161, 52127813, 51975565), in part by the Shanghai Rising-Star Program (20QA1404200), in part by the Shanghai Science and Technology Innovation Action Plan (20dz1206700), and in part by the Joint Fund of Science \& Technology of Liaoning Province and State Key Laboratory of Robotics of China (2020-KF-22-12).}
	
	\bibliographystyle{IEEEtran}
	%Bibliography file should be compiled with BibTex
	\bibliography{mybibfile}
	
\end{document}